\documentclass[aps,prd,reprint,superscriptaddress,nofootinbib,longbibliography]{revtex4-2}

\usepackage{graphicx}
\usepackage{amsmath,amssymb,booktabs,siunitx,hyperref}
\usepackage{hyperref}
\usepackage{xcolor}
\usepackage{booktabs}
\usepackage{tikz}
\usepackage{comment}

\usetikzlibrary{arrows.meta,positioning,fit,calc}

\DeclareSIUnit{\MeV}{MeV}
\DeclareSIUnit{\MeVfm}{MeV.fm}
\DeclareSIUnit{\fm}{fm}
\newcommand{\MeVpc}{\ensuremath{\mathrm{MeV}/c}}

\begin{document}

\title{BeAGLE++: A modern modular framework for $e$A physics at the EIC}
\author{Arjun Kumar}
\email{ arjun.kumar.2@stonybrook.edu}
\affiliation{Center for Frontiers in Nuclear Science (CFNS),
	Stony Brook University, Stony Brook, New York 11794, USA}
	
\author{Abhay Deshpande}
\affiliation{Center for Frontiers in Nuclear Science (CFNS),
	Stony Brook University, Stony Brook, New York 11794, USA}
\affiliation{Physics Department, Brookhaven National Laboratory,
	Upton, New York 11973, USA}
	
		\author{Zhoudunming Tu}
	\affiliation{Center for Frontiers in Nuclear Science (CFNS),
		Stony Brook University, Stony Brook, New York 11794, USA}
	\affiliation{Physics Department, Brookhaven National Laboratory,
		Upton, New York 11973, USA}

\date{\today}

\begin{abstract}
We present the Benchmark eA Generator for LEptoproduction (BeAGLE++), a next-generation, general-purpose Monte Carlo event generator for high-energy lepton–nucleus collisions and the successor to BeAGLE. The framework incorporates a formation-zone intranuclear cascade to describe hadronic final-state interactions (FSIs), followed by statistical de-excitation of the residual nucleus. Compared with the legacy implementation, the BeAGLE++ cascade introduces a discrete time-ordered treatment of hadronic interactions and light-cluster production during intranuclear transport. Comparisons with E665 measurements and legacy BeAGLE calculations show broad agreement for selected observables, while revealing differences in final states sensitive to FSIs. These differences provide opportunities to assess the model dependence of nuclear-breakup predictions and constrain the underlying cascade dynamics. With its modular architecture, BeAGLE++ provides a flexible simulation framework for the Electron–Ion Collider (EIC) and a platform for developing and testing new physics models against existing data and future EIC measurements.

\end{abstract}


\maketitle

\section{Introduction}
\label{sec:intro}
 
Quantum Chromodynamics (QCD) is the fundamental theory of strong 
interactions that govern the dynamics of quarks and gluons across a broad
range of distance and momentum scales. Nuclei provide a unique QCD
environment where partonic coherence, nuclear many-body dynamics and
high-density collective phenomena can be investigated. In particular,
understanding the confinement, including hadronization, gluon saturation, and the production and
propagation of hadrons through nuclear matter remains central to the study
of QCD~\cite{Accardi:2009qv,Accardi:2005jd,Falter:2004uc,Gallmeister:2007an}. The forthcoming Electron--Ion Collider (EIC) in the US will provide an unprecedented access to these phenomena through high-luminosity collisions of polarized electrons with protons and nuclei over a broad range of center-of-mass energies and momentum transfers ($Q^2$)~\cite{Accardi:2012qut,AbdulKhalek:2021gbh}.

A central challenge in lepton--nucleus scattering is to connect the
short-distance lepton--parton interaction to the observed hadronic final
state. Inclusive fixed-target deep-inelastic-scattering (DIS) measurements have established
that the partonic structure of bound nucleons differs from that of free
nucleons, thereby constraining the nuclear initial state \cite{Aubert:1983xm,Gomez:1993ri,Amaudruz:1995tq,Arneodo:1996ru}. Semi-inclusive measurements at HERMES and Jefferson Lab have further demonstrated that
hadron production is modified by propagation through nuclear matter \cite{Falter:2004uc,Gallmeister:2007an,Airapetian:2007vu,Moran:2021pfj}.
Following the hard interaction, the struck system evolves from a colored
partonic configuration toward color-neutral hadrons over formation scales
that can be comparable to, or exceed, the nuclear radius. Depending on the
interaction vertex, in-medium path length, and nuclear density profile, the
evolving system can undergo energy loss, rescattering, absorption, and
secondary particle production. Final-state observables consequently depend on the coupled effects of
nuclear structure, collision geometry, formation times, and interaction
cross sections.

The geometry dependence of such observables provides sensitivity to the
space-time evolution of hadronization and intranuclear transport in cold
nuclear matter~\cite{CLAS:2025nbt}. Although this sensitivity is analogous
in spirit to the geometry dependence of final-state observables in
heavy-ion collisions, the underlying dynamical environment in electron-ion ($eA$)
scattering is distinct. A quantitative interpretation of EIC measurements
therefore requires event generators that can systematically vary the initial
nuclear configuration, final-state interactions (FSIs) and and nuclear de-excitation mechanisms. For a recent review of cold-nuclear-matter effects and approaches for separating initial- and final-state contributions, see
Ref.~\cite{nqcd}. 

A complete description of nuclear DIS requires a treatment of both the
current- and target-fragmentation regions. The current-fragmentation region is associated
primarily with the struck partonic system, whereas the hard interaction also
leaves an excited nuclear remnant. Its response populates the
target-fragmentation region through slow hadrons, knocked-out nucleons, light
nuclear clusters, and residual nuclear fragments \cite{Adams:1995is, Adams:1994yz,Adams:1994zm}. These particles provide
complementary information on energy deposition, intranuclear transport, and
the many-body response of the remnant. Correlations between the current- and
target-fragmentation regions can thus constrain hadronization, collision
geometry, and nuclear breakup simultaneously. Realistic simulations of
final-state particles in both regions, including nuclear-remnant products,
are consequently required to exploit the EIC physics program fully.

\begin{figure*}
	\centering
	\includegraphics[width=0.85\linewidth]{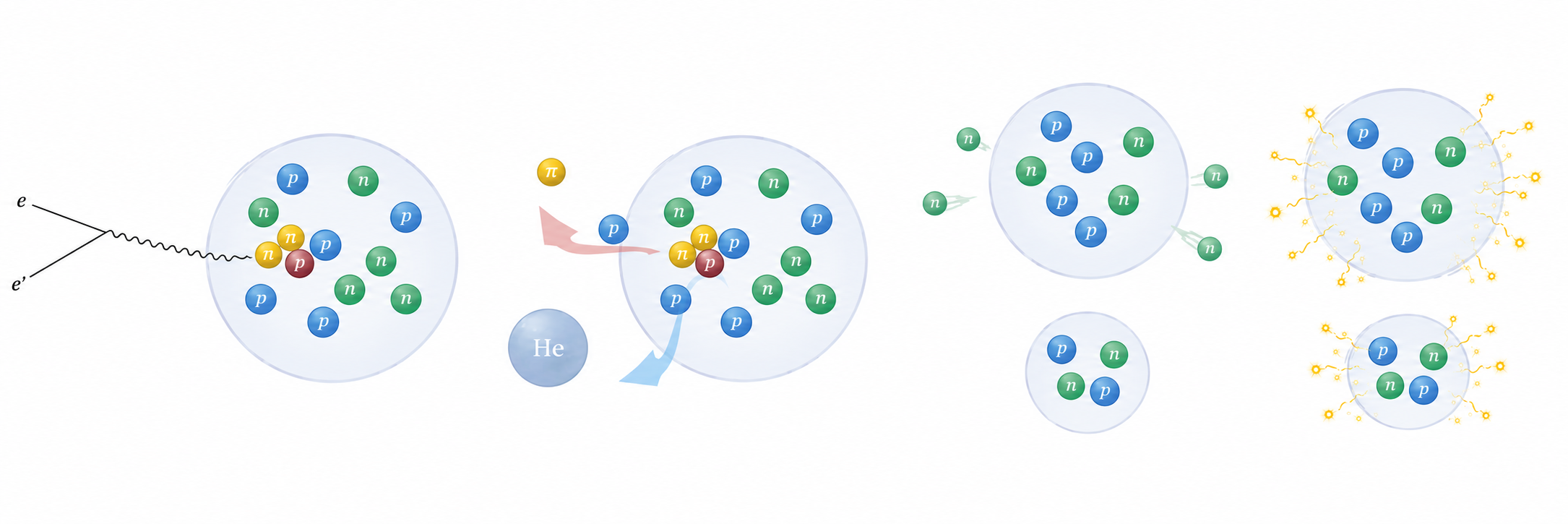}
	
	\caption{Schematic stages of nuclear breakup in $eA$ collisions as modeled
in \textsc{BeAGLE++}: (i) hard scattering of a virtual photon from a bound
nucleon, (ii) final-state interactions of hadrons produced in the primary
interaction within the residual nucleus, including light-cluster emission
during intranuclear transport, and (iii) statistical de-excitation of the
thermalized residual nucleus, leading to the emission of nuclear fragments
and photons.}
	\label{cartoon}

\end{figure*}

Prior to this work, the Benchmark $eA$ Generator for LEptoproduction
(BeAGLE)~\cite{Chang:2022jgo} was the only publicly available event
generator providing a unified description of high-energy lepton--nucleus
interactions with realistic nuclear fragmentation
~\cite{Chang:2021jnu,Jentsch:2021qdp,Tu:2020ymk}. BeAGLE has enabled studies
of collision geometry~\cite{Chang:2022jgo,Hegazy:2024tgt}, spectator
tagging~\cite{Jentsch:2021qdp}, short-range
correlations~\cite{Tu:2020ymk}, diffractive nuclear breakup
~\cite{Chang:2021jnu}, partonic energy loss
~\cite{RoblesGajardo:2022efe}, nuclear excitation
~\cite{Magdy:2025cse}, light-nuclear structure
~\cite{Magdy:2024thf}, rare-isotope production
~\cite{Ddamulira:2026qwo}, and the nuclear response to fast hadrons
~\cite{CLAS:2025nbt}. However, its legacy Fortran implementation makes it
difficult to incorporate, benchmark, and systematically compare new
descriptions of the initial nuclear state, intranuclear transport, and
nuclear de-excitation. In particular, its fixed software structure limits
controlled comparisons between alternative FSI and de-excitation models, and
thus the quantification of model dependence in particle-production and
geometry-sensitive observables.

In this paper, we present BeAGLE++, a modular C++ successor to BeAGLE.
BeAGLE++ retains the established physics framework of the original generator
while introducing a time-ordered intranuclear cascade coupled to nuclear
de-excitation. As illustrated in Fig.~\ref{cartoon}, an $eA$ event evolves
from the primary virtual-photon--nucleon interaction through the propagation
and re-interaction of the produced hadronic system in the nucleus to the
de-excitation of the residual nuclear system. The modular architecture
enables alternative descriptions of the hard interaction, intranuclear
transport, and nuclear de-excitation to be implemented and bench-marked within
a common framework. It thereby permits systematic studies of the model
dependence of FSI-driven predictions under variations of the underlying
intranuclear dynamics.

\begin{table*}[t]
	\caption{\label{tab:legacy}
		Comparison of principal physics modules in BeAGLE++ and legacy BeAGLE model.}
	\begin{ruledtabular}
		\begin{tabular}{lll}
			Component  & \textbf{BeAGLE++} & Legacy BeAGLE \\
			\hline
			Hard process / Hadronization  & \textbf{PYTHIA-8}& PYTHIA-6\\
			Intranuclear cascade   & \textbf{Extended INCL++}& DPMJet \\
			Nuclear De-excitation  & \textbf{ABLA++} & FLUKA \\
			Nuclear geometry & \textbf{BeAGLE++ and INCL++} & BeAGLE and PyQM   \\
			Partonic energy loss  & \textbf{Not yet implemented}& PyQM \\
			Language   &  \textbf{C++ }& Fortran  \\
		\end{tabular}
	\end{ruledtabular}
\end{table*}

The remainder of this paper is organized as follows. Sec.~II provides a detailed overview of the BeAGLE++ event generator, its underlying physics framework and improvements relative to the legacy model. Sec.~III presents the validation of the model against experimental data with comparisons from fixed-target \(\mu\)A data from the E665 experiment. In Sec.~IV, we discuss the generator’s current status and outline key directions for its future development. Finally, in Sec.~V, we summarize the new framework and this work’s main developments.


\section{BeAGLE++ }
\label{sec:framework}

\label{sec:legacy}
BeAGLE++ builds upon the physics foundations established by the legacy BeAGLE event generator. 
It preserves the underlying impulse approximation adopted in the original framework, in which the exchanged virtual photon interacts incoherently with a single bound nucleon. The primary hard interaction and hadronization are simulated with PYTHIA-8~\cite{Bierlich:2022pfr}, while hadronic FSIs are modeled using an extended version of the Li\`ege Intranuclear Cascade model, INCL++~\cite{Boudard:2012cx,Mancusi:2014zka}. The de-excitation of the excited nuclear remnant is subsequently treated with the ABLA++ statistical decay model~\cite{PhysRevC.105.014623,PhysRevLett.130.132501}. These components provide a consistent and up-to-date description of nuclear breakup over a broad range of target nuclei and beam energies. The present implementation focuses on deep-inelastic scattering and does not yet include diffractive final states. Owing to the modular architecture of BeAGLE++, however, dedicated generators of exclusive and diffractive processes, such as Sar$t$re \cite{Toll:2012mb,Toll:2013gda}, can be incorporated straightforwardly in future developments.  The interaction in the model  proceeds through the following sequence of steps:

\begin{enumerate}
	\item An initial bound nucleon is sampled from the nuclear configuration according to the spatial and momentum distributions described in Sec.~\ref{initialstate}. The nucleon species (proton or neutron) determines the elementary lepton--nucleon interaction, which is simulated with PYTHIA~8. The hard scattering is generated at the parton level, with the nucleon cross section modified by the selected nuclear parton distribution function (modular nPDF).
	
	\item The event geometry is reconstructed from the virtual-photon direction and the position of the struck nucleon. Geometrical quantities relevant for subsequent hadronic propagation, including the impact parameter \(b\), nuclear thickness \(T\), and effective path length \(d\), are evaluated.
	
	\item Primary hadrons produced by PYTHIA~8 are assigned a formation length according to the selected formation-time model. Each hadron is then classified as being formed either inside or outside the nuclear volume (see Sec.~\ref{formation}).
	
	\item Hadrons formed inside the nucleus are propagated through the residual nuclear medium using the modified \textsc{INCL++} intranuclear cascade model which results in secondary particles. Hadrons formed outside the nucleus are assumed to escape without experiencing hadronic final-state interactions.
	
	\item Following the intranuclear cascade, the excited residual nucleus is de-excited using the ABLA++ statistical decay model, which describes particle evaporation, Fermi breakup, fission, and $\gamma$-ray de-excitation according to the excitation energy and mass of the nuclear remnant.
\end{enumerate}

For light targets ($A<4$), owing to their small spatial extent, the intranuclear cascade is omitted, across most of the DIS phase space the hadrons  from primary interaction are formed outside the nucleus and thus do not undergo FSIs, in analogy with the legacy BeAGLE treatment of light nuceli FSI suppression. Residual FSI contributions \cite{Cosynn_2011,Strikman:2017koc} may nevertheless remain relevant in limited regions of phase space, these will be incorporated in a future version of the generator. In addition, sampling an off-shell nucleon within the impulse approximation introduces an energy–momentum imbalance that requires careful treatment. A more complete treatment of light nuclei will be incorporated in future versions of the model. 

\subsection{Improvements to legacy BeAGLE Model}
Here we provide a description of the main improvements relative to the legacy model. The description of the nucleus and of the subsequent intranuclear cascade is substantially more physical and consistent. These advances are driven primarily by the sophisticated intranuclear cascade model INCL++, which is coupled to BeAGLE++ and governs the fate of the produced hadrons inside the nucleus. A comparison of the principal physics modules implemented in the legacy version and BeAGLE++ is presented in Table~\ref{tab:legacy}. The key improvements over the legacy model are:

\begin{enumerate}
	\item \textbf{Dynamical nucleus}:
	In BeAGLE++ the nucleus is a dynamical object,  once created, its
	nucleons are classical particles moving in a potential well. In the legacy
	model, nucleons carried Fermi momenta, but their spatial positions remained
	frozen throughout the cascade (a static Glauber/Woods--Saxon snapshot).

	\item \textbf{Propagating participants and Surface dynamics}:
	In BeAGLE++ both participants and spectator nucleons evolve in the potential
	well and can reflect from or transmit through the nuclear surface. In the
	legacy model, soft nucleons below a constant energy threshold are effectively
	removed from the free cascade (trapped, absorbed, or marked as excited), they
	do not bounce and re-scatter.
	
	\item \textbf{Time-ordered cascade versus Generation-ordered cascade}:
	The legacy cascade (\texttt{DT\_FOZOCA}/\texttt{DT\_INUCAS}) is sequential and
	generation-ordered, all free hadrons of generation~$n$ are processed before
	generation~$n+1$. There is no global clock that compares competing
	interactions. This systematically misorders events in which a daughter from an
	early collision should interact before another primary has formed.
	BeAGLE++ uses INCL++'s time-ordered interaction list, so the
	earliest physically allowed event (collision, decay, reflection, or
	transmission) is executed first, and the interaction list is updated after
	each step.
	
	\item \textbf{Formation zone as a spacetime constraint}:
In legacy BeAGLE, formation time is applied independently to each particle, effectively delaying its interaction within the corresponding cascade step without considering the relative timing of other particles. In BeAGLE++/INCL++, formation time is incorporated into the time evolution of the full cascade through a common global clock. Unformed hadrons are prevented from interacting, while already formed particles remain eligible for earlier interactions. This ensures consistent time ordering and a more realistic treatment of formation-zone effects in the multi-particle intranuclear cascade.

	\item \textbf{Richer elementary interaction channels and Higher energy reach}:
	Legacy hadronic reinteractions are limited by the \texttt{HADRIN} module to
	roughly $0.01$--$9\,\mathrm{GeV}/c$, with exclusive channel tables and a
	phenomenological two-nucleon absorption prescription.
	INCL++/BeAGLE++ extends the useful range to
	$\sim 1\,\mathrm{MeV}$--$20\,\mathrm{GeV}$ and includes dynamical channels  \cite{PEDOUX201116,PhysRevC.101.014608, PhysRevC.98.021602}
	absent from the legacy cascade , notably:
	\begin{itemize}
		\item continuous pion production : \\
		$NN\to NN\,x\pi$,
		$\pi N\to N\,(x+1)\pi$, 
        where $x$ denotes the multi-pion production.
		
		\item $\Delta$ dynamics:
		$NN\leftrightarrow N\Delta$,
		$\pi N\leftrightarrow\Delta$,
		$\Delta\to\pi N$
		
		\item $\eta$ and $\omega$ production, scattering, absorption
		
		\item an extended strangeness network, including
		$NN\to NYK\,x\pi$,
		$NN\to NNK\bar{K}$,
		$\pi N\to NK\bar{K}$,
		$NK/N\bar{K}$ multi-pion channels,
		hyperon--nucleon scattering $NY\to NY'$,
		and $\Delta N$-induced strangeness production
		
		\item richer antinucleon annihilation (including $\Xi$ channels and at-rest
		annihilation)
		
		\item in-cascade cluster coalescence (up to $A\le 8$).

	\end{itemize}

\end{enumerate}
\begin{table*}[htbp]
	\centering
	\label{density}
	\small
	\begin{tabular}{llll}
		\toprule
		Regime & \(A\) & Spatial density \(\rho(r)\) & Momentum density \(\pi(p)\) \\
		\midrule
		Deuteron & 2
		& Paris wave function\cite{Lacombe1981}
		& Paris wave function\cite{Lacombe1981} \\
		Light & 3--6
		& Gaussian (\(R_{\mathrm{RMS}}\) from Ref.~\cite{deVries1987})
		& Gaussian \\
		Medium & 6--19
		& Modified harmonic oscillator\cite{deVries1987}
		& Gaussian   \\
		Heavy & \(>19\)
		& Woods--Saxon\cite{Boudard2002}
		& Uniform Fermi sphere \\
		\bottomrule
	\end{tabular}
	\caption{Single-particle spatial and momentum distributions used in BeAGLE++ for different nuclear mass regions.}
\end{table*}
Together, these changes replace the legacy formation-zone geometric filter on
a static nucleus with a chronologically consistent, dynamical intranuclear
cascade, while retaining formation-time suppression for high-energy
secondaries. In the following section, we provide a detailed description of the different stages of the
space--time evolution of an event in BeAGLE++.

\subsection{Initial State and the Hard Interaction }
\label{initialstate}

In BeAGLE++, the initial nuclear configuration is generated by the INCL++ cascade model, which assigns an initial position and momentum to each target nucleon according to mass-dependent spatial and momentum distributions. Prior to sampling, a characteristic Fermi momentum, \(p_F\), is specified for the nucleus. The default INCL++ prescription adopts a constant value
\begin{equation}
	p_F = 1.37\,\hbar c \approx \num{270}\,\MeVpc,
	\qquad \hbar c = \SI{197.328}{\MeVfm}
\end{equation}
which corresponds to the Fermi momentum of saturated nuclear matter. Alternative parameterizations include a light-nucleus lookup table, converted according to
\(p_F=p_{\mathrm{RMS}}\sqrt{5/3}\), and the mass-dependent parameterization of Moniz \textit{et al.}\cite{Moniz1971},
\begin{equation}
	p_F(A)=\alpha-\beta e^{-\gamma A},
\end{equation}
where \(\alpha=\num{259.416}\,\MeVpc\), \(\beta=\num{152.824}\,\MeVpc\), and \(\gamma=9.5157\times10^{-2}\). For asymmetric nuclei, separate proton and neutron Fermi momenta are obtained from
\begin{equation}
	p_{F,p}=p_F\left(\frac{2Z}{A}\right)^{1/3},
	\qquad
	p_{F,n}=p_F\left[2\left(1-\frac{Z}{A}\right)\right]^{1/3},
\end{equation}
after which individual nucleon momenta are sampled from the appropriate momentum distribution. The spatial and momentum distributions employed by \textsc{INCL++} depend on the nuclear mass number, as summarized in Table~\ref{density}. 

For the deuteron (\(A=2\)), nucleon positions and momenta are sampled directly from the Paris \(NN\) potential, including both \(s\)- and \(d\)-wave components\cite{Lacombe1981}. Owing to the two-body nature of the system, only one nucleon is sampled explicitly, while the second is assigned by momentum and position conservation $
	\mathbf{p}_2=-\mathbf{p}_1,
	\qquad
	\mathbf{r}_2=-\mathbf{r}_1 $. For light nuclei (\(3\le A\le6\)), the radial density is described by a Gaussian distribution,

\begin{equation}
	\rho(r)\propto
	\left(\frac{r}{\sigma_r}\right)^2
	\exp\!\left[-\frac12\left(\frac{r}{\sigma_r}\right)^2\right],
	\qquad
	\sigma_r=\frac{R_{\mathrm{RMS}}}{\sqrt3},
\end{equation}

where the rms radii are taken from electron-scattering measurements compiled in Ref.~\cite{deVries1987}. The momentum distribution is Gaussian with

\begin{equation}
	\sigma_p=\frac{p_{\mathrm{RMS}}}{\sqrt3},
	\qquad
	p_{\mathrm{RMS}}=p_F\sqrt{\frac35},
\end{equation}

or from the tabulated light-nucleus values used in the \texttt{constant-light} option. For medium-mass nuclei (\(6\le A\le19\)), nucleon positions are sampled from a modified harmonic oscillator density,

\begin{equation}
	\rho(r)\propto
	r^2
	\left[1+R_0\left(\frac{r}{a}\right)^2\right]
	\exp\!\left[-\left(\frac{r}{a}\right)^2\right],
\end{equation}

with parameters determined from the charge-density analysis of Ref.~\cite{deVries1987}. The momentum distribution remains Gaussian, using the corresponding Fermi momentum. For heavy nuclei (\(A>19\)), the spatial distribution is described by a Woods--Saxon density,

\begin{equation}
	\rho(r)=
	\frac{\rho_0}
	{1+\exp[(r-R_0)/a]},
\end{equation}

where

\begin{align}
	R_0 &= (2.745\times10^{-4}A+1.063)\,A^{1/3}\ \mathrm{fm},\\
	a   &= 1.63\times10^{-4}A+0.510\ \mathrm{fm},
\end{align}
The corresponding momentum distribution is a uniform Fermi sphere,

\begin{equation}
	f(p)\,dp=
	\frac{3p^2}{p_F^3}\,dp,
	\qquad
	0\le p<p_F.
\end{equation}

The model further provides an optional position--momentum (\(r\)--\(p\)) correlation, whereby nucleons with larger momenta are preferentially localized near the nuclear surface rather than uniformly throughout the nuclear volume. Details of the correlation model and its implementation are given in Ref.~\cite{Boudard2002}.

BeAGLE++ presently neglects the effects of nucleon Fermi motion in the calculation of the primary deep-inelastic scattering (DIS) cross section. Consequently, the DIS kinematic variables, such as Bjorken \(x\) and the four-momentum transfer squared \(Q^2\), are generated directly by PYTHIA-8 and are not modified by the initial momentum of the bound nucleon. Since the primary lepton--nucleon interaction in PYTHIA-8 assumes an on-shell nucleon with the free nucleon mass, directly assigning a nonzero initial Fermi momentum would generally violate energy--momentum conservation at the primary interaction vertex. Within the present implementation, this kinematic mismatch is absorbed by the residual nuclear system. After the primary interaction, the excitation energy and recoil momentum of the nuclear remnant are adjusted such that the total four-momentum of the complete electron--nucleus event is conserved. This practical approximation was also used in the legacy BeAGLE framework and helps maintain a consistent description of the subsequent intranuclear cascade and statistical de-excitation.

\subsection{Formation zone Intranuclear Cascade}
\label{formation}
Following the formation-zone approach adopted in the legacy BeAGLE framework~\cite{Chang:2022jgo,Roesler:2000he}, each primary hadron produced during fragmentation is assigned a formation time, during which it is assumed not to undergo hadronic interactions with the surrounding nuclear medium. The average proper formation time is parameterized as

\begin{equation}
	\langle\tau\rangle
	=
	\tau_0
	\frac{m^2}{m^2+p_T^2},
	\label{eq:tau}
\end{equation}

where \(m\) and \(p_T\) denote the hadron mass and transverse momentum respectively, and \(\tau_0\) is a free model parameter. The actual proper formation time is sampled from an exponential distribution with mean \(\langle\tau\rangle\), after which it is Lorentz-boosted to the laboratory frame to determine the corresponding formation length. The parameter \(\tau_0\) is constrained primarily by measurements of forward neutron production in \(\mu\)--Pb collisions~\cite{E665:1995utr}.A hadron is classified as formed inside the nucleus if its formation point lies within the nuclear density distribution, or equivalently if its formation length is smaller than the distance to the nuclear surface along its trajectory. Only hadrons satisfying this criterion are propagated through the residual nucleus and are therefore subject to hadronic FSIs.

The propagation of formed hadrons is performed using an extended version of the Li\`ege Intranuclear Cascade model, INCL++. The cascade describes the transport of hadrons through the nuclear medium via successive binary hadron--nucleon collisions while enforcing Pauli blocking for final-state nucleons \cite{Boudard:2012cx,Mancusi:2014zka}. The model incorporates elastic and inelastic scattering, charge-exchange reactions, resonance production and decay, as well as nuclear mean-field effects where applicable. Owing to its extensive validation against hadron--nucleus reaction data over a broad range of projectile species and energies, INCL++ is expected to provide a reliable description of hadronic final-state interactions in electron--nucleus collisions. BeAGLE++ does not use INCL++ as a stand-alone beam--nucleus cascade. Instead, slow hadrons from photon-nucleon interaction are injected at their formation vertices. This is the principal extension relative to stock INCL++ usage, the cascade is seeded by an ensemble of  hadrons formed inside the nucleus rather than a single projectile. A formation time based sampling is also done for the subsequent generation of hadrons produced during the cascade.

\subsection{Residual Nucleus Breakup}
\label{sec:abla}

After the cascade, the residual prefragment is de-excited with the ABLA++
statistical model. ABLA describes particle evaporation, fission where applicable, and related
decay channels of hot nuclei~\cite{Kelic:2008zza}. Light remnants are  treated with Fermi breakup according to INCL/ABLA configuration limits. The resulting nucleons, light clusters, and photons are appended to the event record with a de-excitation origin tag. A detailed technical description of the Monte Carlo framework, including the event-generation architecture, steering mechanism, and runcard interface, is provided in Ref.~\cite{technical}.

Since the production of forward neutrons is dominated by the intranuclear cascade and the subsequent statistical de-excitation of the residual nucleus, the average neutron multiplicity provides a sensitive constraint on the hadron formation time. In BeAGLE++, the formation-time parameter \(\tau_0\) is therefore calibrated using the forward neutron multiplicity measured in \(\mu\)--Pb deep-inelastic scattering by the Fermilab E665 experiment~\cite{E665:1995utr}. The current implementation of BeAGLE++ describes only inelastic lepton--nucleus interactions and does not include coherent diffractive scattering, for which the nucleus remains intact and no evaporation neutrons are produced. The E665 neutron multiplicity measurements, however, contain contributions from both inclusive and diffractive processes. Consequently, the simulated neutron multiplicity is corrected by the coherent event fraction,$	f \equiv
N_{\rm coherent}/N_{\rm total}$ such that the measured average neutron multiplicity is expressed as
\begin{equation}
	\langle N_n\rangle_{\rm E665}
	=
	(1-f)\,
	\langle N_n\rangle_{\rm BeAGLE++},
	\label{equation:E665}
\end{equation}
where coherent events are assumed to contribute negligibly to the forward neutron yield. 
\begin{figure}
	
	\centering
	\includegraphics[width=0.92\linewidth]{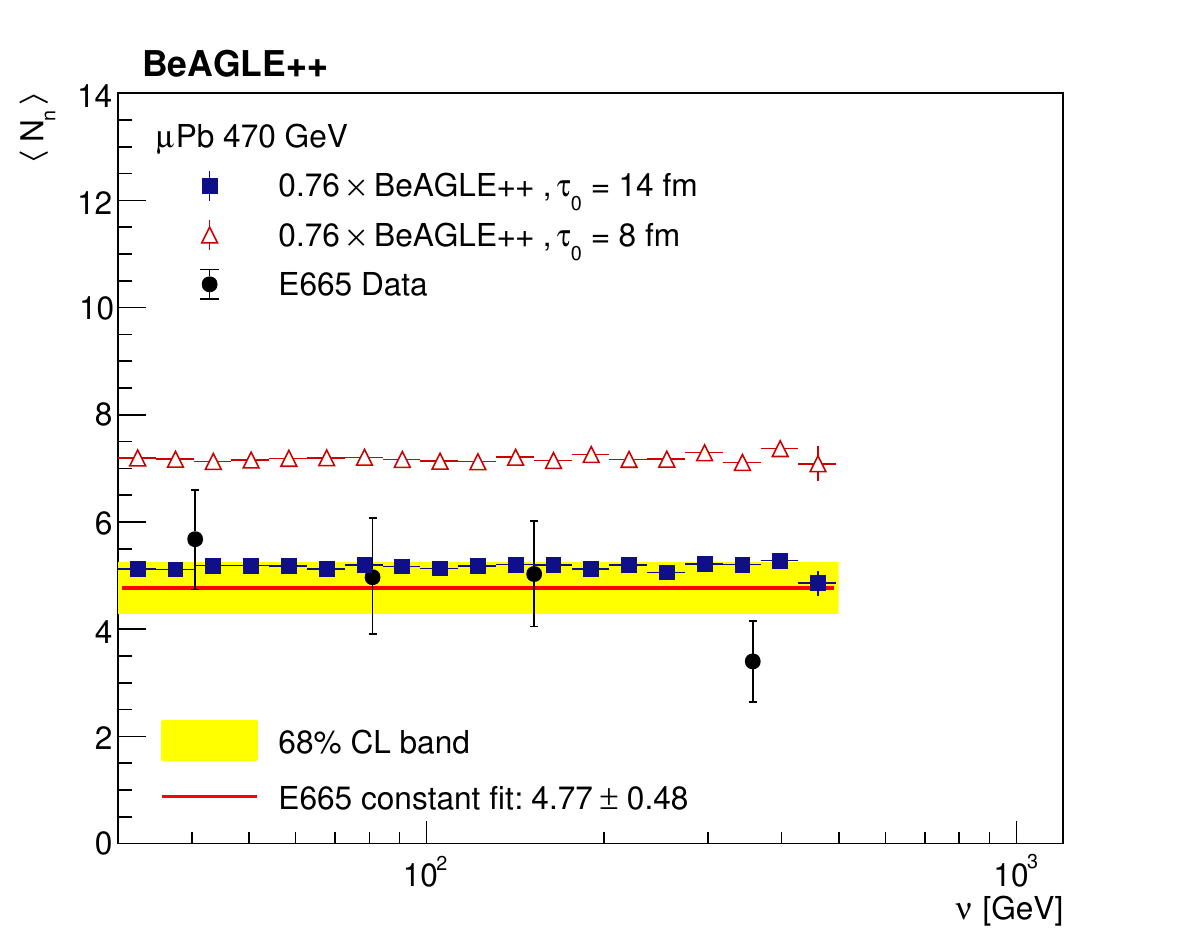}\\
	
	\caption{\label{neutronmulitplicity}Average neutron multiplicity $\langle N_n\rangle$ versus photon energy $\nu$ in $\mu\mathrm{Pb}$ collisions at $470~\mathrm{GeV}$ using BeAGLE++,  Black points: E665 data, yellow band:  $\pm1\sigma$ constant fit uncertainty.}
\end{figure}

\begin{figure*}[htbp]
	\centering
	\setlength{\tabcolsep}{1pt}
	\begin{tabular}{@{}c@{\hspace{15pt}}c@{}}
		\includegraphics[width=0.82\columnwidth]{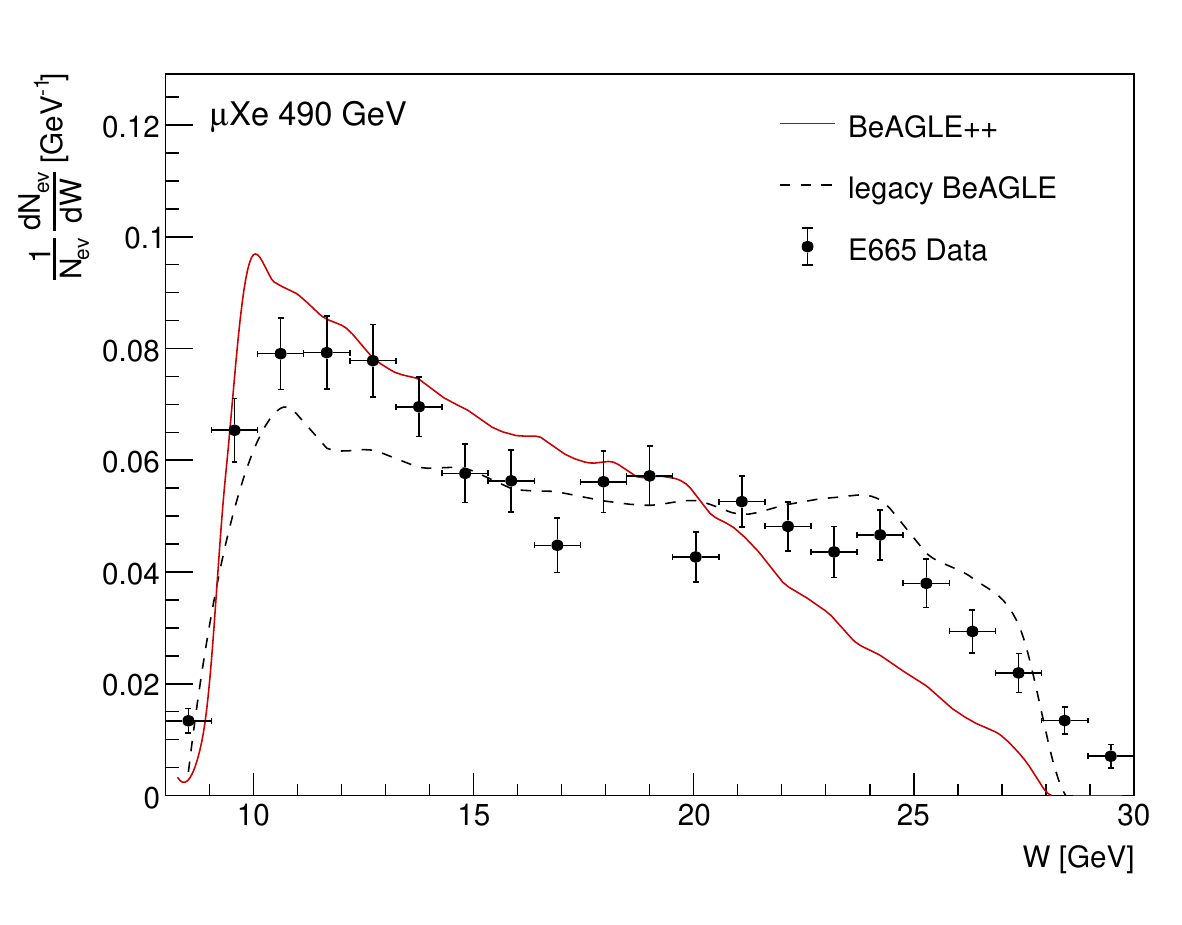} &
		\includegraphics[width=0.82\columnwidth]{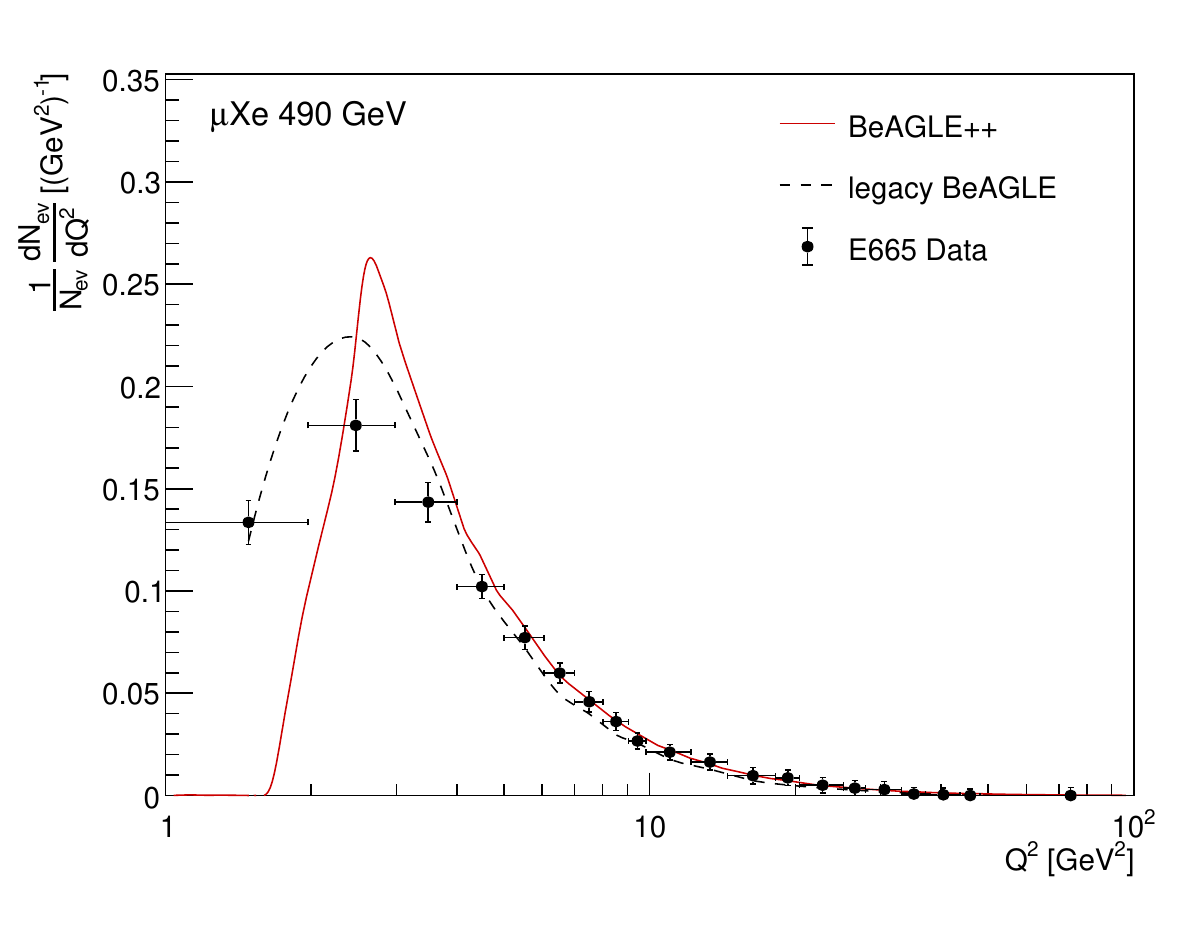} \\[-0.4em]
		(a) & (b) \\[0.4em]
		\includegraphics[width=0.82\columnwidth]{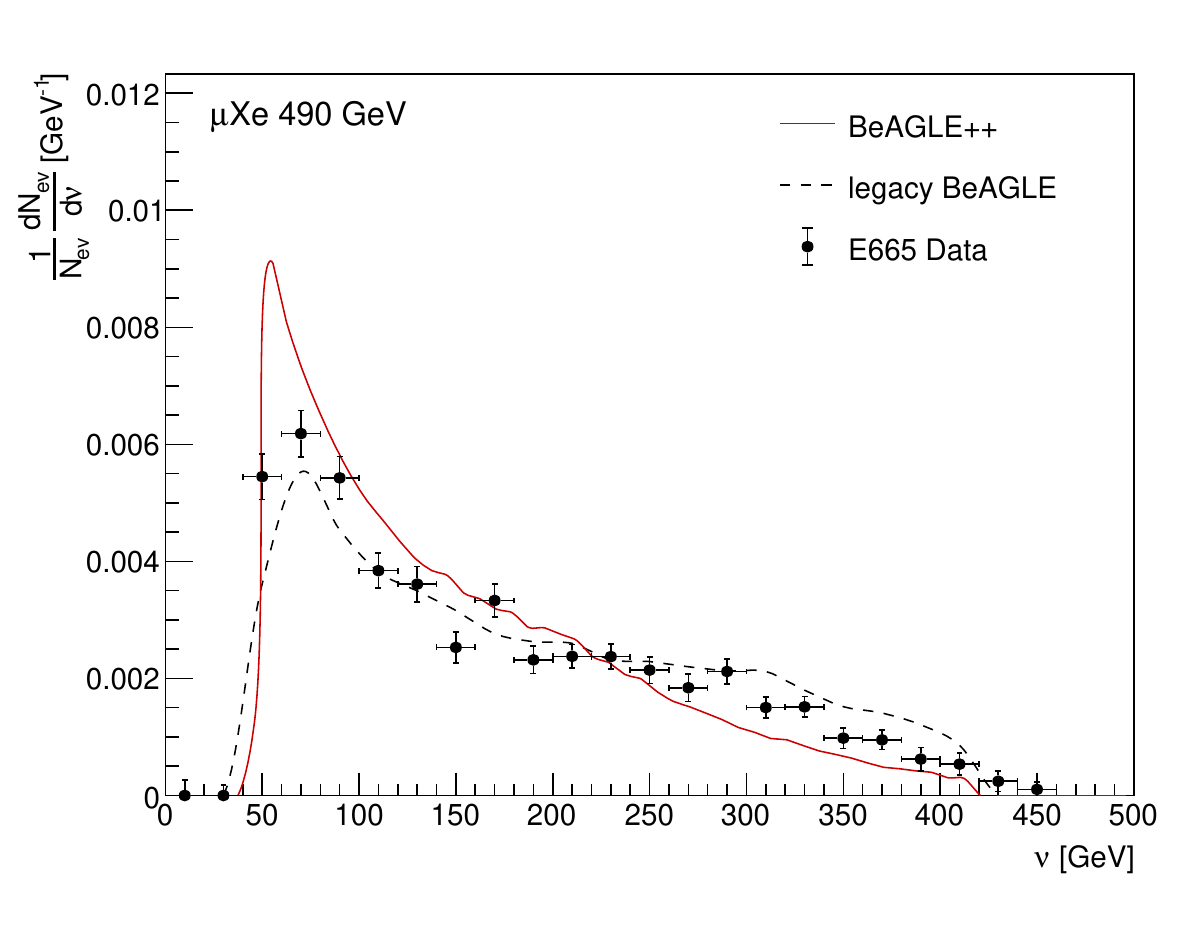} &
		\includegraphics[width=0.82\columnwidth]{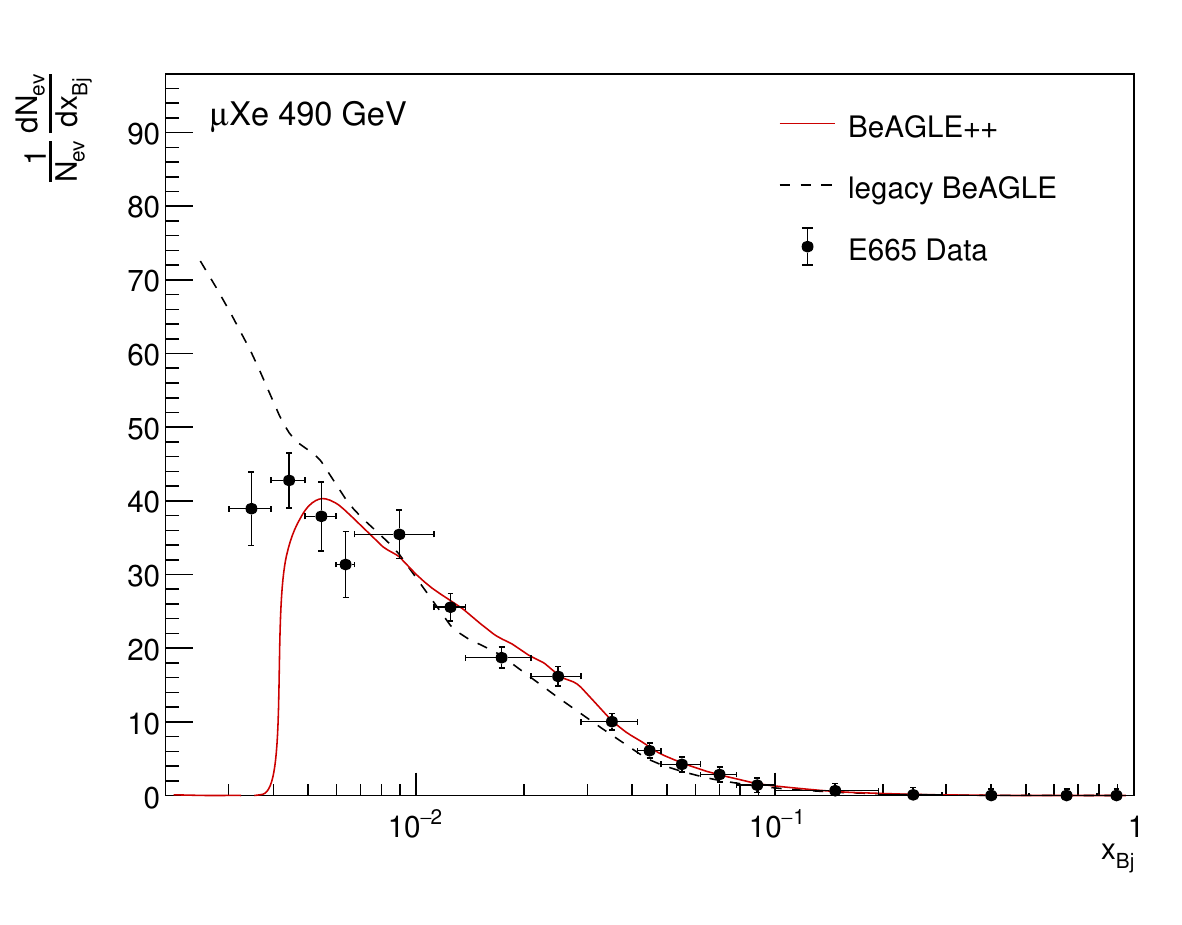} \\[-0.4em]
		(c) & (d)
	\end{tabular}
	\caption{\label{kindist}Normalised kinematic distributions in $\mu\mathrm{Xe}$ collisions at $490~\mathrm{GeV}$ as functions of (a)~$W$, (b)~$Q^{2}$, (c)~$\nu$, and (d)~$x_{\mathrm{Bj}}$. Red solid curves show BeAGLE++ events, black dashed curves show legacy BeAGLE, and black markers denote E665 data.}
\end{figure*}
Figure~\ref{neutronmulitplicity}  compares the average neutron multiplicity, \(\langle N_n\rangle\), as a function of the virtual-photon energy \(\nu\), measured by E665 with the corresponding BeAGLE++ predictions for two different values of the formation-time parameter \(\tau_0\), assuming a representative coherent fraction of \(f=24\%\). A constant fit to the experimental data yields $	\langle N_n\rangle_{\rm E665}=4.77\pm0.48,$ where the uncertainty corresponds to one standard deviation. Under the assumption \(f=24\%\), the best agreement with the measured neutron multiplicity is obtained for $\tau_0 = 14~{\rm fm}$ constrained to lie within the $\pm1 \sigma$ uncertainity of the constant fit to the data. The extracted value of the parameter $\tau_0$ is larger than that used in the legacy BeAGLE implementation ($\tau_0 = 10~\mathrm{fm}$). This increase is primarily attributable to the substantially revised and more comprehensive treatment of final-state interactions (FSIs) in the current model. 

Unlike the legacy implementation, the updated FSI framework incorporates a larger set of interaction channels and surface dynamics as discussed previously, resulting in a more realistic description of intranuclear hadron transport. The inclusion of these additional channels enhances energy deposition within the residual nucleus, leading to greater nuclear excitation and, consequently, increased neutron emission during the de-excitation process. As a result, the fit favors a larger value of $\tau_0$, which compensates for the enhanced FSI dynamics and provides improved agreement with the neutron multiplicty E665 data as depicted in Fig. \ref{neutronmulitplicity}. 

For the remainder of this paper, the extracted value of the formation-time parameter, $\tau_0 = 14~\mathrm{fm}$, is used as the default in all simulations and is employed for comparisons with the legacy BeAGLE model and the E665 experimental data.

\section{Data and MC Comparison}
\label{validation}
In this section, the predictions of the BeAGLE++  are systematically compared with available experimental measurements. Measurements of charged-hadron production in muon--xenon ($\mu$Xe) deep-inelastic scattering reported by E665~\cite{E665:1993trt} and predictions from legacy BeAGLE are used as the primary experimental references for evaluating the performance of the BeAGLE++ model.

\begin{figure*}[htbp]
	\centering
	\setlength{\tabcolsep}{1pt}
	\begin{tabular}{@{}c@{\hspace{15pt}}c@{}}
		\includegraphics[width=1.5\columnwidth]{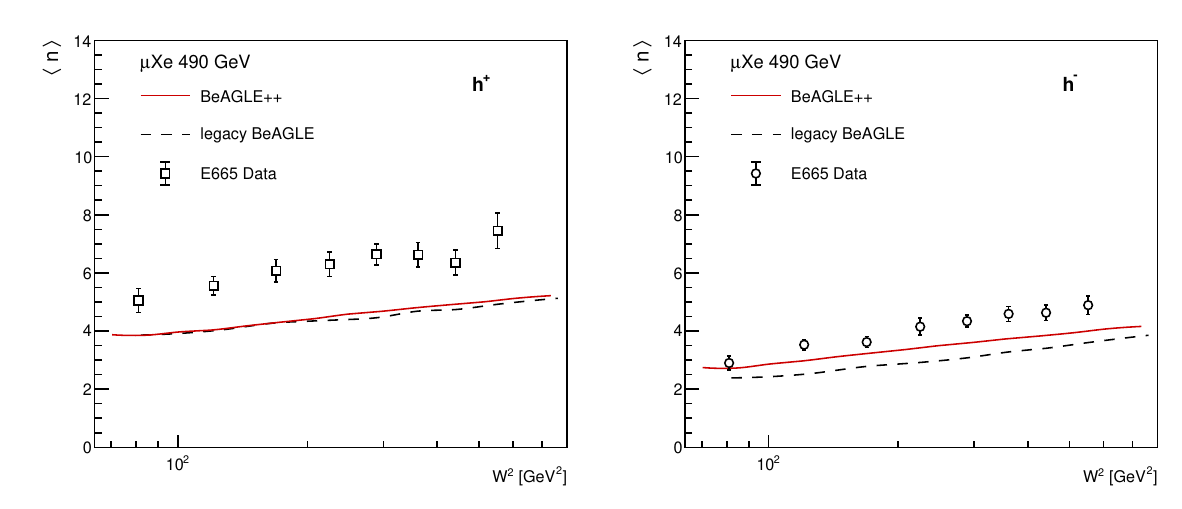} \\
		(a)  \\[0.4em]
		\includegraphics[width=1.5\columnwidth]{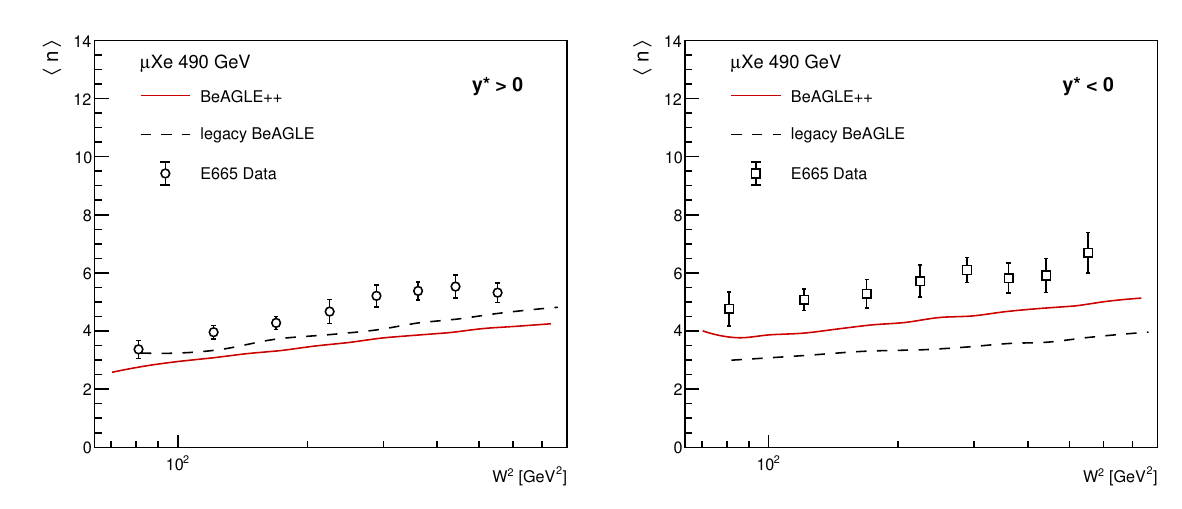} \\
		(b) 
	\end{tabular}
	\caption{\label{hadmult}Average charged hadron multiplicity distributions as a function of $W^{2}$ in $\mu\mathrm{Xe}$ interactions at $490~\mathrm{GeV}$: positively and negatively charged hadrons (top row), and positive and negative $y^{*}$ (bottom row). BeAGLE++ simulations (solid red curves) are compared with legacy BeAGLE (dashed black curves) and E665 data (open black markers) from~\cite{E665:1993trt}.}
\end{figure*}

Figure~\ref{kindist} compares the distributions of the DIS kinematic variables $W$, $Q^{2}$, $\nu$, and $x_{\rm Bj}$ predicted by BeAGLE++ with the corresponding measurements from the E665 $\mu$Xe experiment~\cite{E665:1993trt} and the legacy BeAGLE predictions. To reproduce the experimental event selection, the simulated events were required to satisfy the same kinematic criteria as those applied in the data analysis: $\theta > 3.5~\mathrm{mrad}$, $Q^{2} > 1~\mathrm{GeV}^{2}$, $8 < W < 30~\mathrm{GeV}$, and $0.1 < \nu/E_{\mu} < 0.85$. The BeAGLE++ predictions are shown as red solid curves and the legacy version predictions are black dashed lines, while the acceptance-corrected E665 measurements~\cite{E665:1993trt} are represented by black solid circles.  

Overall, the BeAGLE++ model provides a satisfactory description of the measured DIS kinematic distributions over most of the accessible phase space. Noticeable discrepancies, however, are observed in the low-$x_{\rm Bj}$ and low-$Q^{2}$ regions, where the simulation tends to deviate from the experimental measurements. These discrepancies are primarily attributable to differences in the modeling of the underlying hard lepton--nucleon interaction at low $Q^{2}$. The current BeAGLE++ implementation employs PYTHIA-8 as the primary event generator, whose description of the low-$Q^{2}$ transition region differs from that of PYTHIA-6 used in the legacy BeAGLE framework. 

The PYTHIA-6-based implementation was empirically found to provide better agreement with the E665 data in this kinematic regime. Consequently, the BeAGLE++ predictions do not exactly reproduce the legacy BeAGLE results, particularly in the low-$Q^{2}$ and correspondingly low-$x_{\rm Bj}$ regions. The description of the low-$Q^{2}$ region in PYTHIA-8 is currently under active development. Ongoing improvements to the modeling of this kinematic regime are expected to enhance the agreement between the BeAGLE++ predictions and experimental data in future releases.

Figure~\ref{hadmult}(a) show the average multiplicity, $\bar{n}$, of positively and negatively charged hadrons produced in $\mu$Xe collisions with 490 GeV muons. BeAGLE++ simulations are compared with the E665 data and the legacy BeAGLE simulations. The description of positively charged hadron production is found to be essentially unchanged with respect to the legacy BeAGLE implementation. For negatively charged hadrons, however, BeAGLE++ provides a modest but systematic improvement in agreement with the experimental data.
The dependence of the  BeAGLE++ predictions on the formation-time parameter, $\tau_{0}$, is not presented here, as its qualitative behavior is consistent with that observed in the legacy BeAGLE implementation and is not the primary focus of this study. The legacy BeAGLE results shown in this comparison correspond to the default value of the formation-time parameter, $\tau_{0}=10~\mathrm{fm}$.

Figure~\ref{hadmult}(b) presents the average charged-particle multiplicity as a function of the center-of-mass rapidity, $y^{\ast}$, obtained from the BeAGLE simulations and compared with the E665 measurements. Positive values of $y^{\ast}$ correspond to the current-fragmentation region, which is dominated by the hadronization of the struck parton, whereas negative values of $y^{\ast}$ define the target-fragmentation region, where particle production is strongly influenced by the nuclear remnant and FSIs. Both the legacy BeAGLE and BeAGLE++ implementations underestimate the measured charged-particle multiplicity. Nevertheless, BeAGLE++ provides a noticeably improved description of the target-fragmentation region relative to the legacy version. This improvement is primarily attributed to the enhanced FSIs framework implemented in BeAGLE++.

\begin{figure*}[htbp]
	\centering

	\includegraphics[width=1.65\columnwidth]{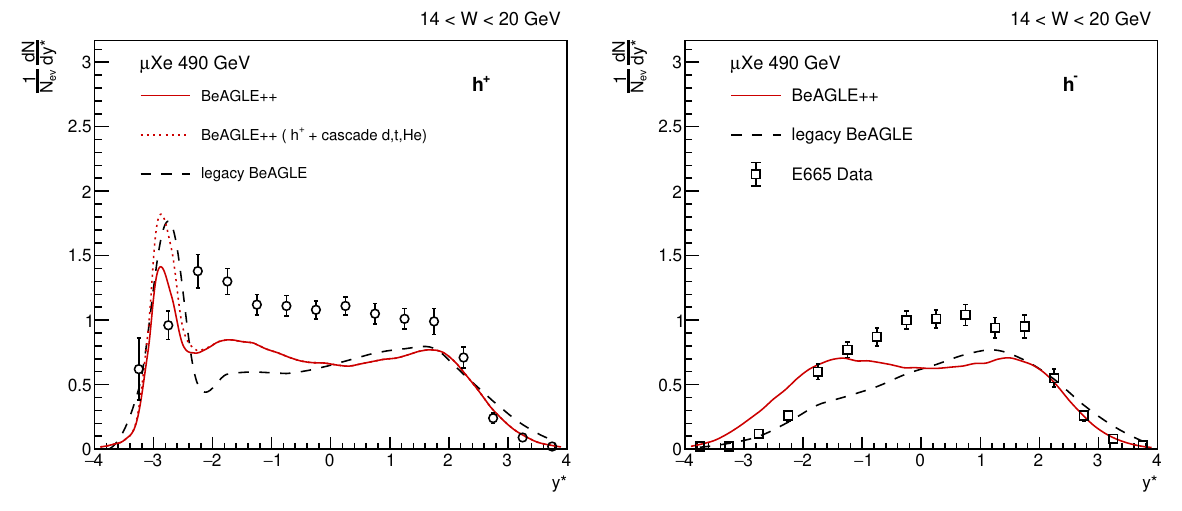} \\
	
	\caption{\label{diffmult}Centre-of-mass rapidity distributions $y^{\ast}$ for positively (left) and negatively (right) charged hadrons in the bin $14 < W < 20~\mathrm{GeV}$ for $\mu\mathrm{Xe}$ interactions at $490~\mathrm{GeV}$. BeAGLE++ simulations (red solid curves) are compared with legacy BeAGLE (black dashed curves) and E665 data (open black markers)~\cite{E665:1993trt}. Red dotted curve (left) depicts the sum of postively charged hadrons and light clusters ($A<4$) emitted during the cascade phase in BeAGLE++.}
\end{figure*}

Figure~\ref{diffmult} shows the normalized center-of-mass rapidity ($y^{*}$) distributions of positively and negatively charged hadrons in $490~\mathrm{GeV}$ $\mu$Xe deep-inelastic scattering, comparing the E665 measurements with the predictions of  BeAGLE++. The model reproduces the overall features of the measured rapidity spectra, however, it systematically underpredicts hadron production in the forward ($y^{*}>0$) current-fragmentation region. In the backward ($y^{*} < 0$) target-fragmentation region, BeAGLE++ (red solid curve) predicts a peak that is slightly displaced relative to the data and less sharply peaked than the legacy BeAGLE prediction. This difference arises because the legacy version does not include light-cluster emission, so a larger fraction of protons are knocked out in the cascade while in BeAGLE++ some outgoing protons are bound into light clusters, thereby reducing the net positive charge carried by free charged hadrons. 

To verify this interpretation, we also show the sum of charged hadrons and light clusters ($A < 4$) as the red dotted curve from BeAGLE++ which reproduces the peak height of the legacy BeAGLE prediction. We note that the E665 data for positively charged hadrons also favour the peak height obtained when light-cluster emission is included.A detailed discussion of the shifted peak position is given in \cite{Chang:2022jgo}, and similar arguments apply to BeAGLE++. Relative to the legacy version, BeAGLE++ provides an improved description of negatively charged hadrons, but slightly overestimates the multiplicity towards the large negative values of rapidity. 

The remaining deficit in particle production within the current-fragmentation region is likely a consequence of the impulse approximation employed in both the legacy BeAGLE and BeAGLE++ frameworks, where the virtual photon interacts with a single bound nucleon. At moderate to low Bjorken-$x$, this approximation is expected to become inadequate, as coherent scattering and multi-nucleon interaction mechanisms contribute significantly to the final-state particle yield. A more complete description would require extending the interaction model to include multiple wounded nucleons within a Glauber-based formalism which would result in more soft particle production. Such developments lie beyond the scope of the present work and will be investigated in future studies.

 Predictions for $e$Ag collisions at the EIC Early Science configuration $9\times115~\mathrm{GeV}$ are provided in  Appendix~\ref{appendixA} illustrating the species and origin dependence of far-forward particle yields relevant for ZDC and forward-proton acceptance studies. A detailed description of the collision geometry with the new sophisticated cascade treatment and the new de-excitation module is also provided in Appendix~\ref{appendixB}, demonstrating the robustness and consistency of the methodology.

\section{Outlook}
\label{sec:outlook}

Several studies with the legacy BeAGLE generator have already established the model's applicability to lepton--nucleus physics at high energies. In the present framework, the emphasis has been on inclusive particle production. Addressing more differential observables, and simulating semi-inclusive and exclusive final states, requires additional physics input that will extend the model's physics reach.

In its current form, the framework already provides a powerful tool for studying particle production in the target-fragmentation region and for constraining the collision geometry. The collision geometry determined in BeAGLE++ is consistent with that obtained using the legacy BeAGLE model. An impact-parameter handle would be crucial for isolating saturation and final-state effects that are geometry driven, nevertheless, the model in its present form does not yet address the full set of physics questions that will be explored at future lepton--ion facilities. Below we outline several priority avenues for model development and physics exploration.

\begin{itemize}
	\item  \textbf{Diffraction}: Inclusion of coherent diffractive scattering on the whole nucleus and the nucleon inside the nucleus, which is absent from the present inelastic-only implementation and affects the interpretation of forward-neutron yields through the coherent fraction $f$ in Eq.~\eqref{equation:E665}. This would also improve the modelling of nuclear breakup in incoherent diffractive configurations, where the nucleus dissociates.
	
	\item  \textbf{Energy loss}:A thorough implementation of partonic energy loss is required to elucidate the space--time evolution of hadronisation in semi-inclusive measurements.
	This effect is closely tied to the collision geometry, is currently missing from the framework, and constitutes a natural next step for development.

	\item  \textbf{Saturation}: Implementing saturation physics and exploring centrality-dependent observables, in connection with a Glauber-based multi-nucleon interaction scheme, would enhance soft-particle production in the current-fragmentation region, where both BeAGLE++ and the legacy generator currently underpredict the E665 multiplicities. In addition, exploratory study using the most up-to-date accelerator scheme at the EIC, e.g., using a 9 GeV electron beam scattering with $\sim 110$ GeV heavy nuclei beam, can be studied in this new model. 
	
	\item  \textbf{Tagged-DIS}: The tagged-DIS programme requires a careful treatment of residual final-state interactions in light nuclei. These interactions are expected to be important for nuclear modifications at high $x$ at future colliders. A fully consistent treatment of nucleon Fermi motion, based on realistic nuclear spectral functions, will improve the description of light ions.
	
	\item  \textbf{Initial-State}: Improvements in the modeling of the nuclear initial state are also needed, encompassing short-range correlations, cluster formation, nuclear deformation, and neutron-skin effects, together with a clearer assessment of the role of light clusters and forward charged particles in centrality determination and of correlations between current- and target-fragmentation particles, including their mass dependence.

\item  \textbf{QED effects}: Incorporate QED radiative corrections in both the initial and final state of the lepton beam. These effects are strongly coupled to the target species and demand a sophisticated, channel-dependent treatment.

	\item  \textbf{De-excitation models}: Integration of alternative nuclear de-excitation models such as \textsc{GEMINI++, FLUKA} to enable a systematic assessment of the model dependence of nuclear breakup.

	\item  \textbf{Validation}: Further validation against forthcoming tagged target-fragmentation data from Jefferson Lab, and against ultra-peripheral collision measurements of the nuclear breakup chain at RHIC and the LHC, will be essential ahead of EIC running.
\end{itemize}

BeAGLE++ is developed primarily for the EIC, including evolving detector
requirements and the prospect of a second interaction
region~\cite{Kim:2026ytt,Chang:2025pgi}. It is also applicable to
fixed-target electron--nucleus measurements at Jefferson Lab
~\cite{BURKERT2020163419} and to future lepton--hadron facilities, including
the Large Hadron--Electron Collider (LHeC)~\cite{Agostini_2021} and the
Electron--Ion Collider in China (EicC)~\cite{Anderle_2021}. BeAGLE++ thus
provides a framework for developing and testing models of lepton--nucleus
interactions over a broad kinematic range.

\section{Summary}
\label{sec:summary}

We have introduced the newly developed framework of BeAGLE++ and demonstrated its application to the study of particle production in lepton--nucleus collisions.  While most event generators focus primarily on the current-fragmentation region, BeAGLE++ provides a detailed description of particle production in the target-fragmentation region, including a full treatment of nuclear breakup. BeAGLE++ is a modular, next-generation C++ Monte Carlo generator for high-energy lepton--nucleus collisions that succeeds the legacy BeAGLE framework.

The hard process and hadronization are generated with PYTHIA-8, hadrons formed inside the nucleus propagate through an extended time-ordered INCL++ cascade; and the excited remnant is de-excited with ABLA++. Relative to the legacy DPMJet/FLUKA chain, the new cascade employs a dynamical nuclear potential, a richer set of elementary channels, and in-cascade light-cluster coalescence, while retaining a formation-zone suppression.  Comparisons with E665 $\mu$Xe data and with legacy BeAGLE show satisfactory agreement for DIS kinematic distributions and charged-hadron multiplicities, with  improvements in the target-fragmentation region attributable to the revised FSI treatment. Differences associated with light-cluster emission, absent in the legacy model during the cascade process, are visible in the backward rapidity spectra of positively charged hadrons. Together, these results establish BeAGLE++ as a maintainable and extensible platform for nuclear-breakup phenomenology at the EIC and related lepton--nucleus facilities.

\begin{acknowledgments}
We thank the authors of the original BeAGLE project: E.~Aschenauer, M.~D.~Baker, J.~H.~Lee, and L.~Zheng for establishing the physics program that motivates the present rewrite, and the developers of PYTHIA~8, INCL++, and ABLA for making their codes publicly available. We are grateful to W.~Chang for providing predictions from legacy BeAGLE simulations. We also thank Thomas Ullrich, the local ePIC Group at Brookhaven National Laboratory and the Center for Frontiers in Nuclear Science (CFNS) members at Stony~Brook University for valuable discussions and feedback on the model. A.Kumar thanks J.~L.~R.~S\'anchez and J.~C.~David for discussions concerning the INCL++ code. The work of A. Kumar and A. Deshpande is supported by Center for Frontiers in Nuclear Science. Z. Tu is supported by the U.S. Department of Energy under Contract No. DE-SC001270 and BNL Laboratory Directed Research and Development (LDRD) 27-045

\end{acknowledgments}

\bibliography{bibliography}
\clearpage

\appendix

Here we present additional comparisons between the updated BeAGLE++ framework and its legacy counterpart. These comparisons extend the analyses already reported in the main paper and are shown explicitly for completeness. The results further corroborate the robustness of final-state-interaction (FSI) driven predictions for particles produced in the target-fragmentation region.

\section{Net charge particle yield}
\label{appendixA}
\begin{figure}[htbp]
	
	\centering
	\includegraphics[width=0.88\linewidth]{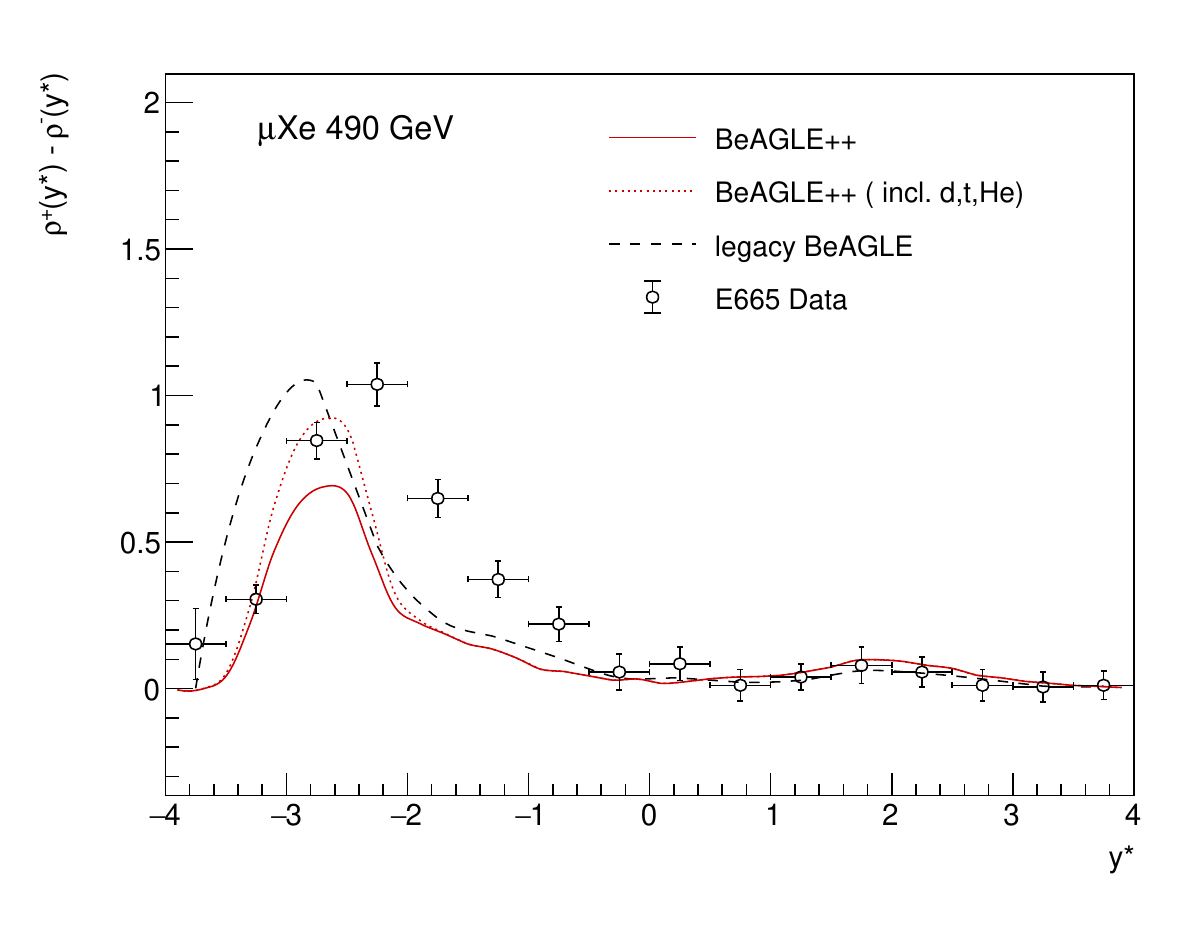}\\
	
	\caption{\label{diffchghadrons}Normalised centre-of-mass rapidity ($y^{\ast}$) distributions of the hadronic net charge	in $\mu\mathrm{Xe}$ collisions with a $490~\mathrm{GeV}$ muon beam.}
\end{figure}

As pointed out in earlier study with legacy BeAGLE simulations, the rapidity dependence of the net charged-particle yield is known to be particularly sensitive to final-state interactions (FSIs), as it enhances the contribution from secondary hadrons produced through intranuclear cascading while suppressing the dominant contribution from the primary hard scattering process. Figure~\ref{diffchghadrons} presents the normalized center-of-mass rapidity ($y^{*}$) distributions of the net charged-particle density,
$\rho^{+}(y^{*})-\rho^{-}(y^{*})$, for $\mu$Xe collisions. Here, the normalized charged-particle rapidity density, $\rho^{\pm}(y^{*})$, is defined as
\begin{equation}
	\rho^{\pm}(y^{*})=\frac{1}{N_{\mathrm{ev}}}\,
	\frac{dN^{\pm}}{dy^{*}},
\end{equation}

where $N_{\mathrm{ev}}$ denotes the total number of selected events and $N^{\pm}$ is the number of positively or negatively charged hadrons, respectively. The qualitative behaviour of both the legacy and new models is consistent with the data, but neither accurately reproduces the measured shape.
This discrepancy can be traced to differences between the simulations and measured rapidity spectra of positively and negatively charged hadrons. Both the models place the peak at a slightly displaced position relative to the data.

\begin{figure*}[htbp]
	\centering
	\setlength{\tabcolsep}{1pt}
	\begin{tabular}{@{}c@{\hspace{15pt}}c@{}}
		\includegraphics[width=0.9\columnwidth]{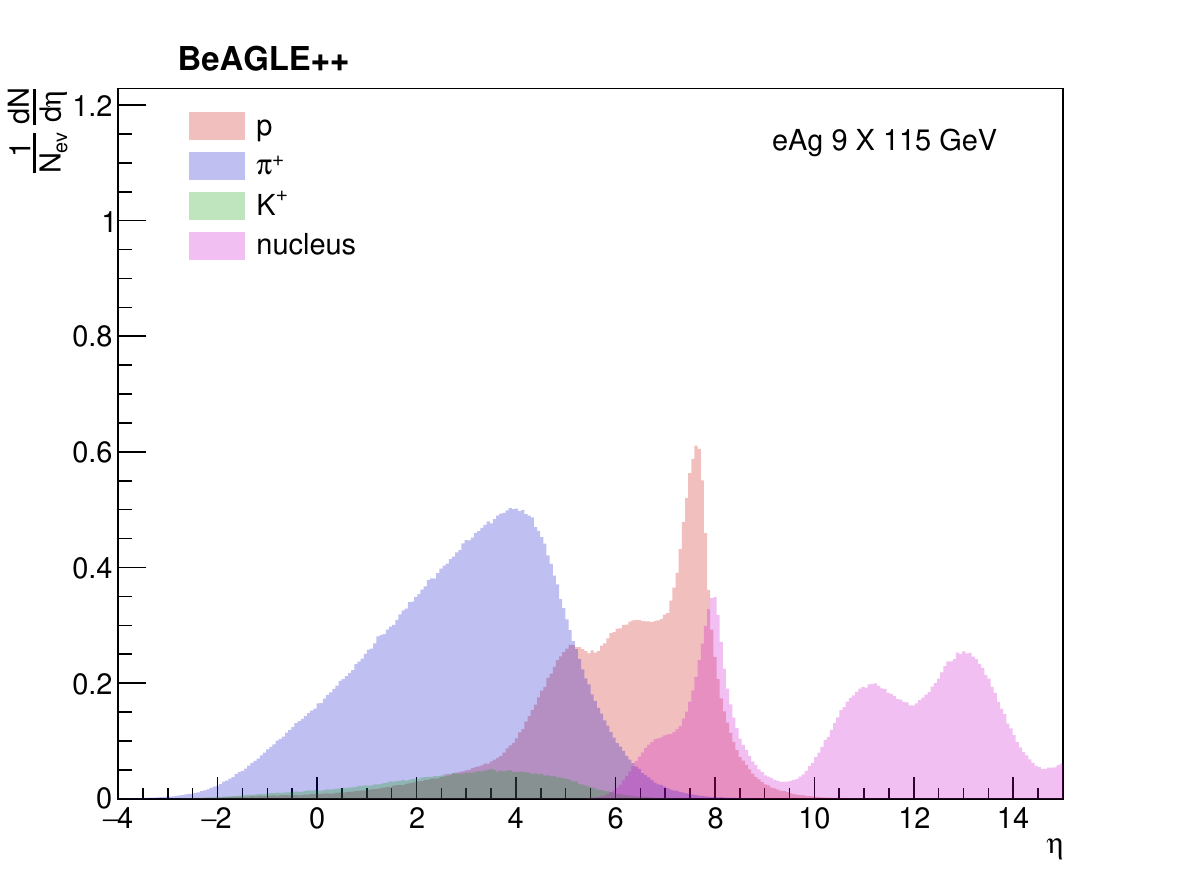} &
		\includegraphics[width=0.9\columnwidth]{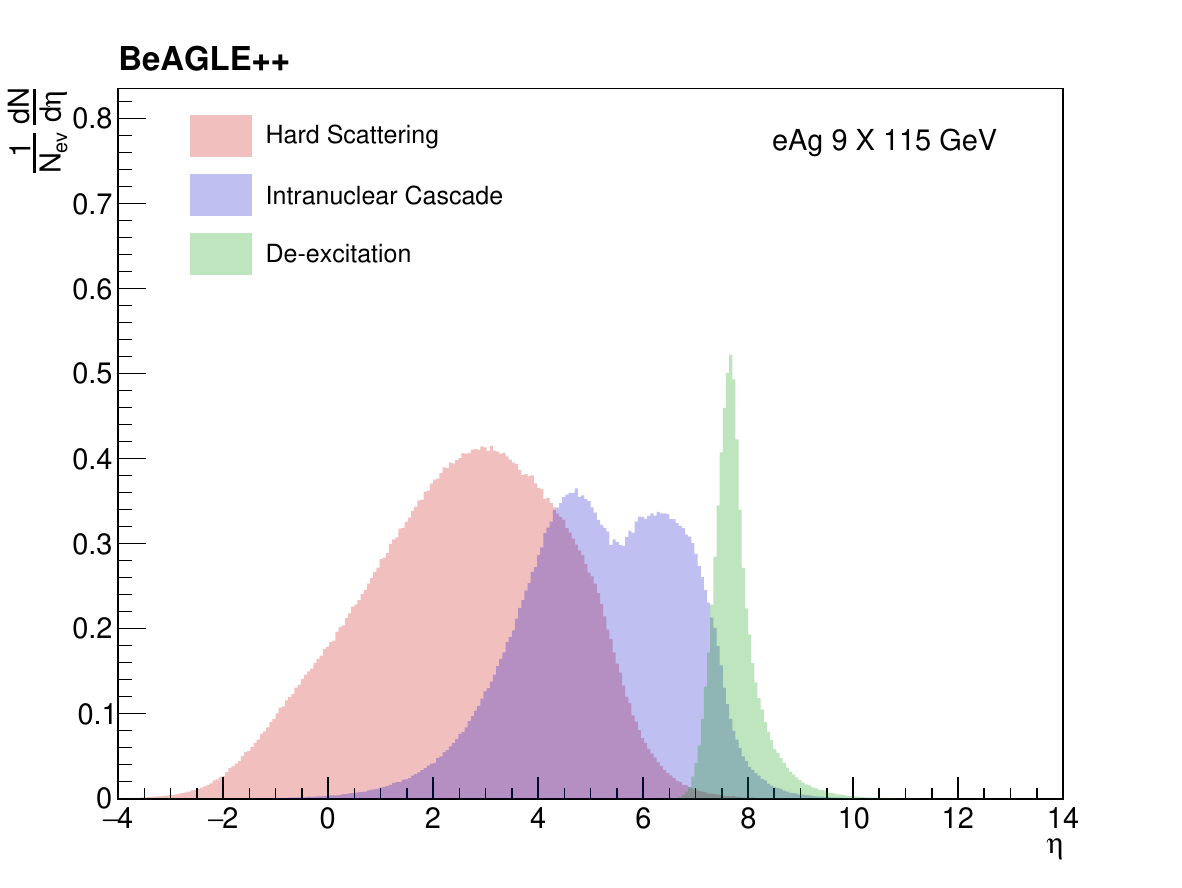} \\[-0.4em]
		(a) & (b) \\[0.4em]
	\end{tabular}
	\caption{\label{EICpred}Left: Normalised pseudorapidity distributions of positively charged particles and nuclear remnants in $e\mathrm{Ag}$ collisions at $9\times115~\mathrm{GeV}$ from BeAGLE++. Different colours correspond to different particle species. Right: Normalised pseudorapidity distributions of positively charged particles classified by their production stage in the BeAGLE++ interaction steps.}
\end{figure*}

Figure~\ref{EICpred} (a) shows the predictions for normalized pseudorapidity ($\eta$) distributions of positively charged particles and nuclear remnants after breakup for an $e$Ag sample corresponding to the EIC Early Science running configuration at $9\times 115\,\mathrm{GeV}$, as simulated with BeAGLE++. Different colors denote distinct particle species. Pions and kaons are produced predominantly in the hard scattering and populate a broad interval $-4\lesssim\eta\lesssim 7$, which lies within the acceptance of the anticipated general-purpose EIC detector. Protons extend over a much wider range, $0\lesssim\eta\lesssim 10$; a substantial fraction of the far-forward proton phase space at $\eta\gtrsim 4.5$ is covered by the suite of dedicated forward proton detectors---the B0 tracker, the Off-Momentum Detector, and the Roman Pots~\cite{AbdulKhalek:2021gbh}. Light nuclear clusters arise primarily from the intranuclear cascade and subsequent nuclear de-excitation, and are concentrated at large positive rapidities ($6\lesssim\eta\lesssim 10$). Heavier nuclear fragments, by contrast, are generated in the final stage of the BeAGLE++ event chain and appear at still more forward pseudorapidities ($\eta\gtrsim 10$), occupying two distinct kinematic domains. Reconstructing these nuclear remnants remains a major experimental challenge and constitutes an active area of development for the future EIC, to be addressed through optimization of the far-forward instrumentation and the design of the second interaction region~\cite{AbdulKhalek:2021gbh}. Figure~\ref{EICpred}\,(b) shows the normalized pseudorapidity ($\eta$) distributions of charged particles, differentially tagged according to their production stage in the BeAGLE++ event chain. This origin classification, separating, for example, particles generated in the hard scattering from those emerging in the intranuclear cascade, nuclear de-excitation illustrates the model's ability to resolve particle production throughout the space--time evolution of each event. Such differential tagging underscores the realism of the lepton--nucleus interaction treatment implemented in BeAGLE++ and the consistency of its nuclear-breakup modeling. This will play a crucial role in forthcoming studies with model to investigate correlations between fast (hard-scattered) and slow (cascade) particles, aimed at probing the nuclear response to fast hadrons in $e$A collisions at the EIC.

\section{Collision geometry}
\label{appendixB}

Legacy BeAGLE demonstrated that forward neutrons correlate with nuclear geometry~\cite{Chang:2022jgo}. BeAGLE++ stores per-event $b$ (impact parameter), $T$ (thickness), and $d$ (effective path length), described belwo, enabling similar studies of centrality proxies at the EIC. This serves a dual purpose, first, it allows us to test and cross-check the methods established for centrality determination in $e$A collisions: until now, no complete model of nuclear breakup other than legacy BeAGLE was available for such a validation and establishing the robustness of method. Second, the new framework can be used to make predictions for early EIC science and for the planned run-matrix configurations. Following the legacy model, the effective interaction length, \(d\), is defined as the density-weighted path length traversed by the virtual photon from the photon--nucleon interaction point to the nuclear surface along the photon propagation direction. It is given by

\begin{equation}
	d(b,z_0)=\int_{z_0}^{+\infty}\mathrm{d}z\,
	\frac{\rho(b,z)}{\rho_0},
	\label{equation:d}
\end{equation}
where \(z_0\) denotes the longitudinal coordinate of the struck nucleon, \(b\) is the impact parameter, \(\rho(b,z)\) is the local nuclear density, and \(\rho_0\) is the central nuclear density. The normalization by \(\rho_0\) ensures that \(d\) has units of length.

The collision geometry can also be characterized by the scaled nuclear thickness function,
\begin{equation}
	\frac{T(b)}{\rho_0}
	=\int_{-\infty}^{+\infty}\mathrm{d}z\,
	\frac{\rho(b,z)}{\rho_0},
	\label{equation:Tb}
\end{equation}
\begin{figure*}[htbp]
	\centering
	\setlength{\tabcolsep}{1pt}
	\begin{tabular}{@{}c@{\hspace{15pt}}c@{}}
		\includegraphics[width=0.88\columnwidth]{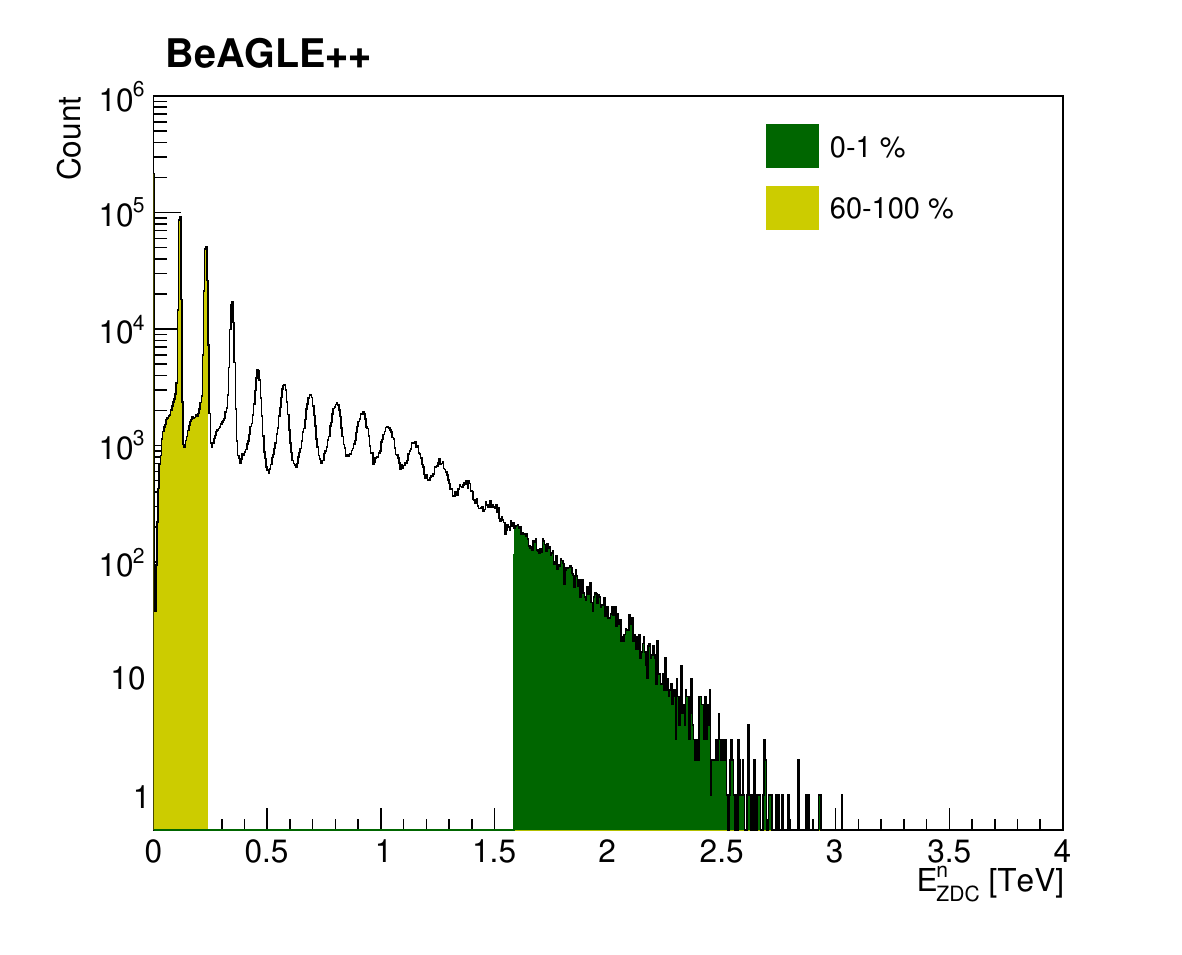} &
		\includegraphics[width=0.88\columnwidth]{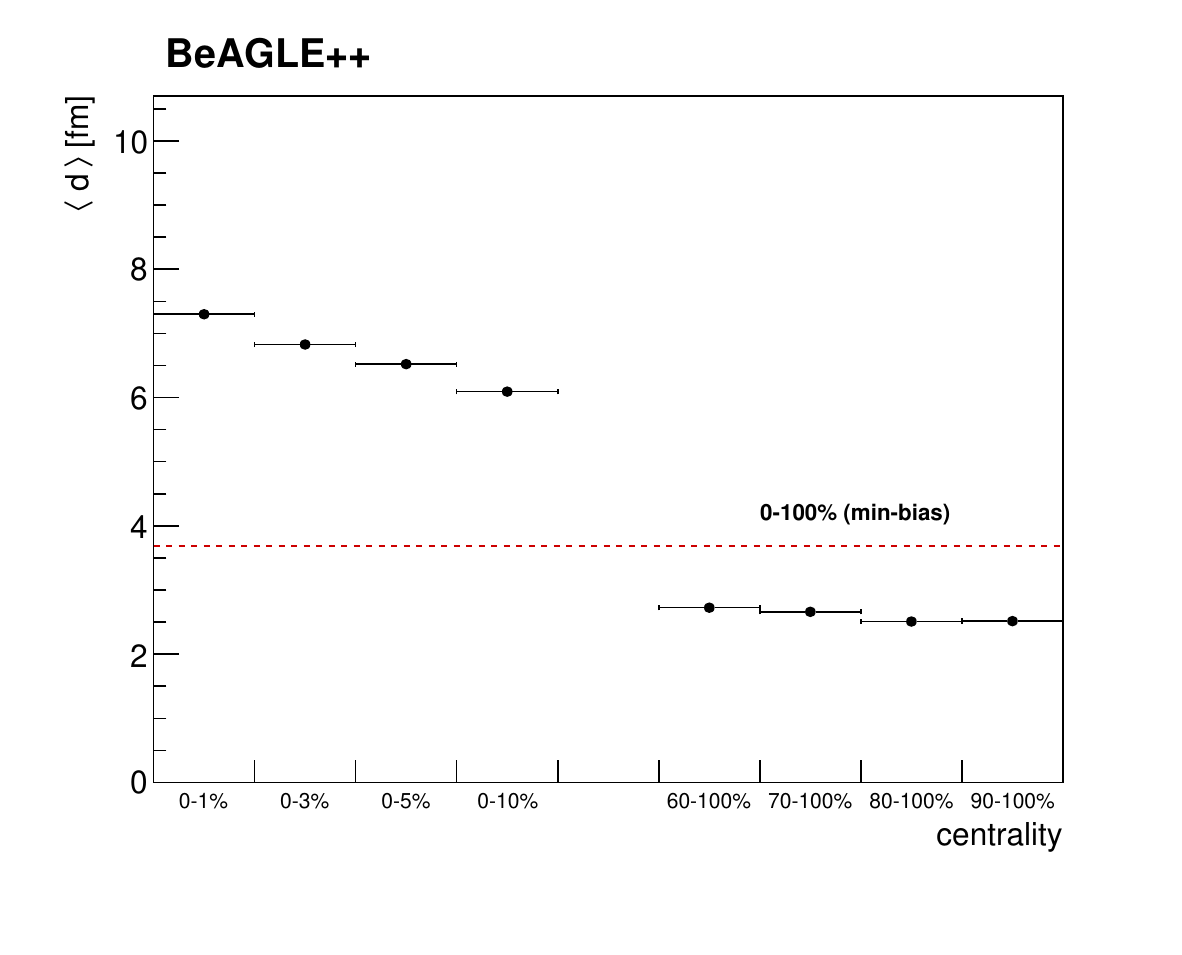} \\[-0.4em]
		(a) & (b) \\[0.4em]
	\end{tabular}
	\caption{\label{EZDC}(a)~Centrality selection based on the energy deposited in the ZDC.
		(b)~Distributions of effective path length $\langle d\rangle$ for central selections $0$--$1\%$, $0$--$3\%$, $0$--$5\%$, and $0$--$10\%$, and for peripheral selections $60$--$100\%$, $70$--$100\%$, $80$--$100\%$, and $90$--$100\%$.}
\end{figure*}

\begin{figure*}[htbp]
	\centering
	\setlength{\tabcolsep}{1pt}
	\begin{tabular}{@{}c@{\hspace{25pt}}c@{}}
		\includegraphics[width=0.65\columnwidth]{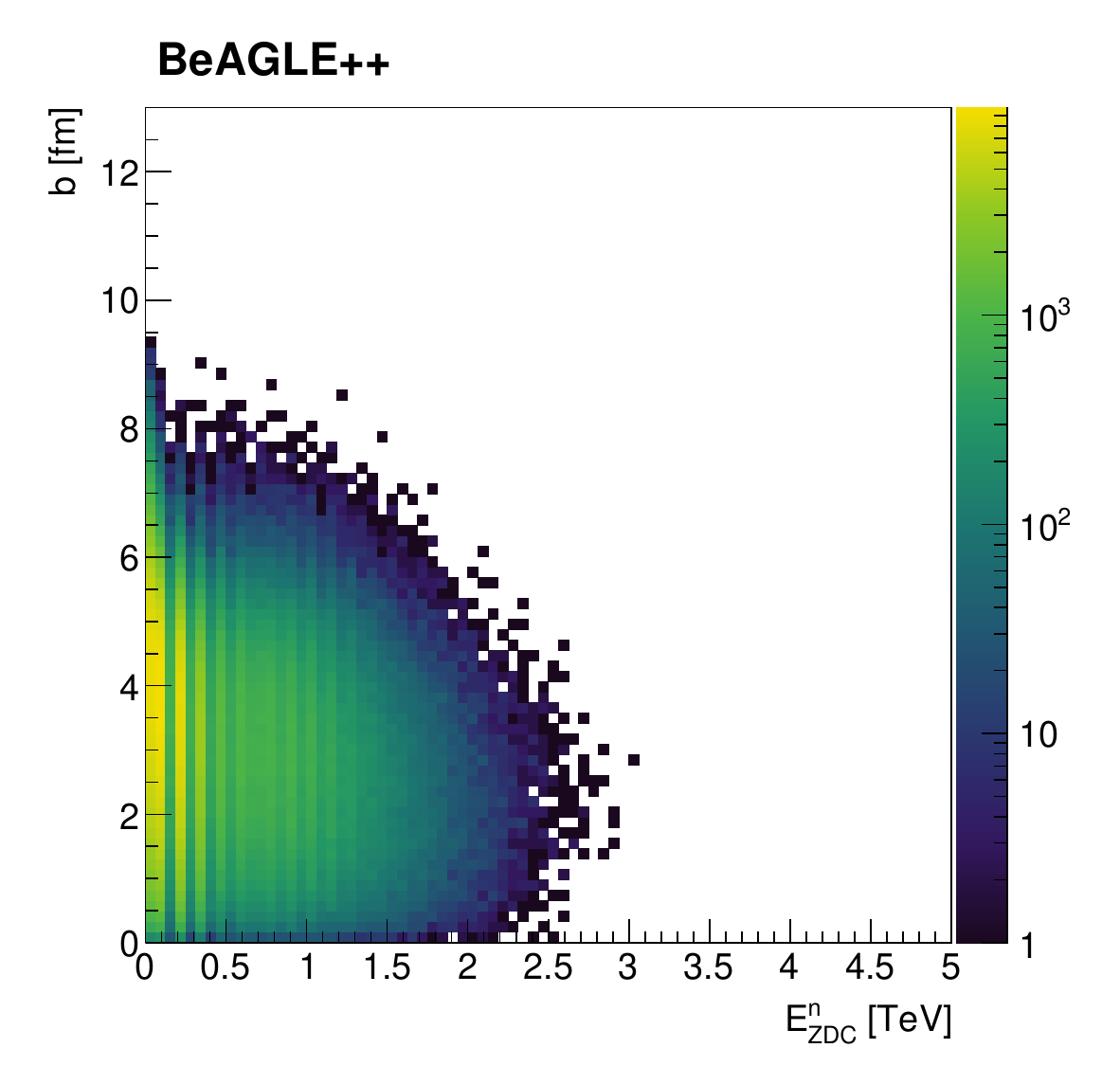} &
		\includegraphics[width=0.65\columnwidth]{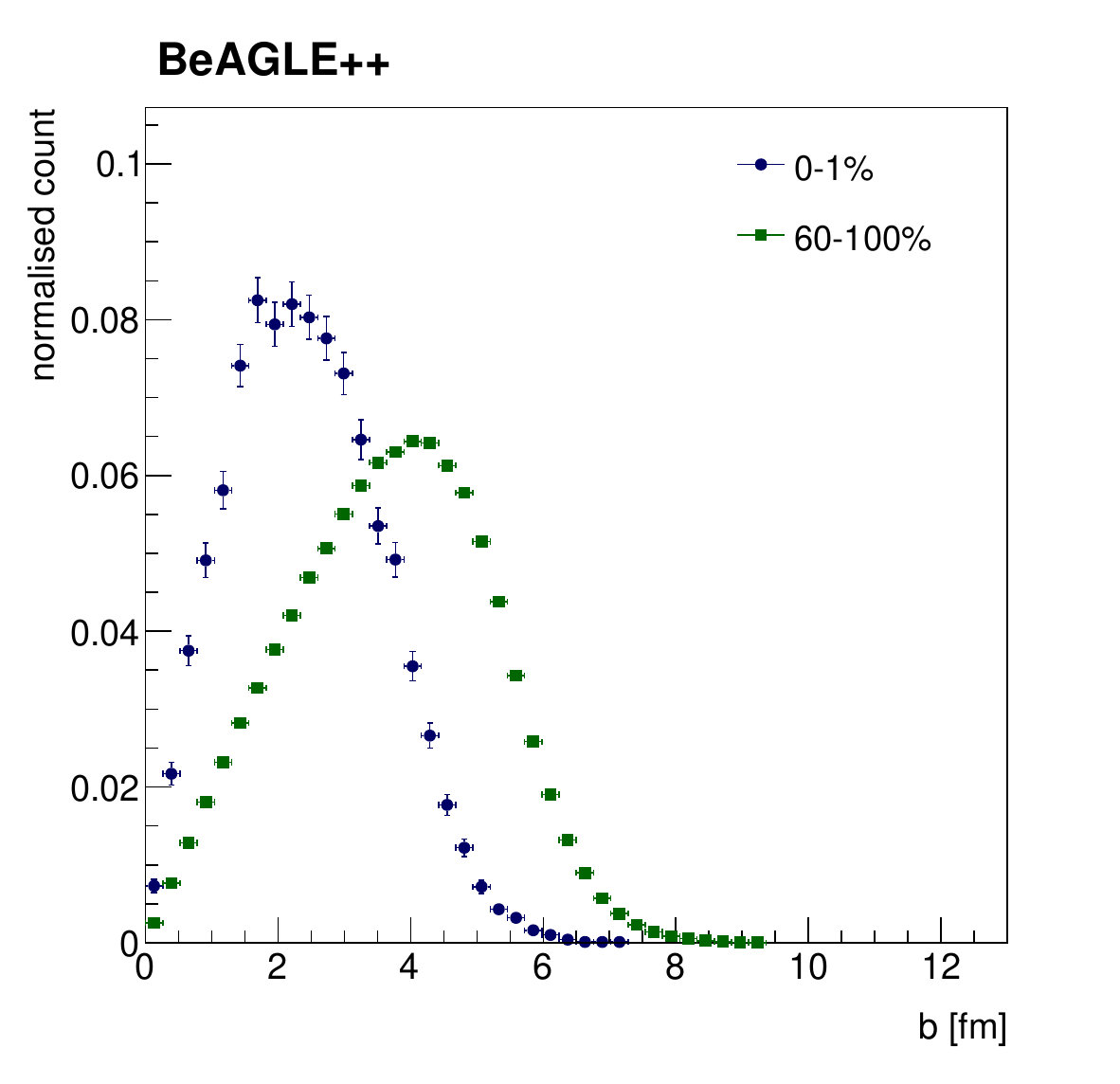} \\[-0.4em]
		(a) & (b) \\[0.4em]
		\includegraphics[width=0.65\columnwidth]{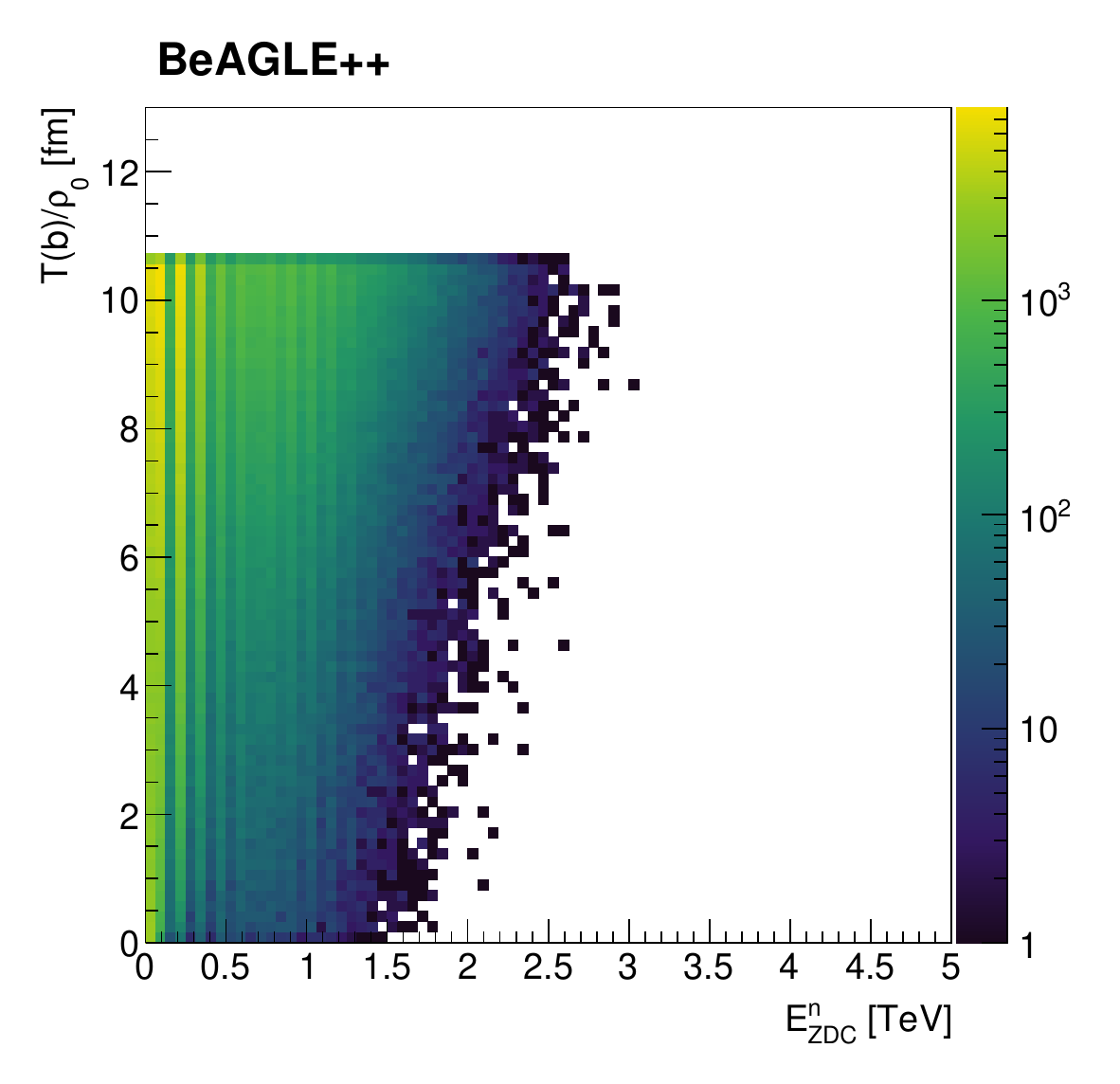} &
		\includegraphics[width=0.65\columnwidth]{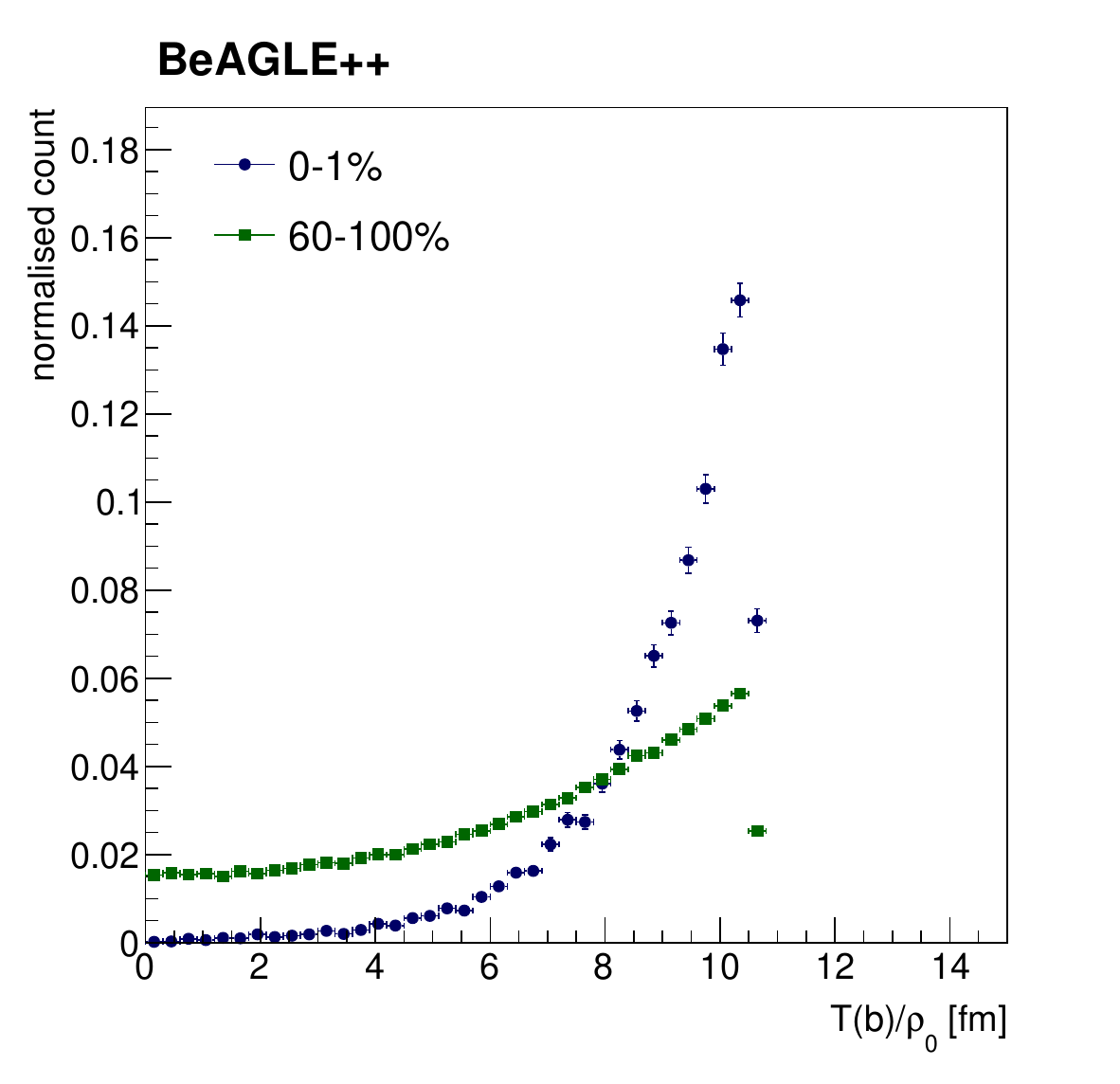} \\[-0.4em]
		(c) & (d)
	\end{tabular}
	
	\caption{	\label{centrality}(a)~Correlation between the ZDC deposited energy and the impact parameter $b$. (b)~Impact-parameter distributions for central and peripheral collisions. (c)~Correlation between the ZDC deposited energy and the nuclear thickness $T(b)/\rho_{0}$. (d)~Distributions of $T(b)/\rho_{0}$ for central and peripheral collisions..}
\end{figure*}

which represents the density-weighted longitudinal extent of the nucleus at a given impact parameter and is likewise expressed in units of fm. Unlike the effective interaction length, which depends on the interaction point \(z_0\), the scaled thickness function depends only on the impact parameter and provides a global characterization of the nuclear geometry. As such, it serves as a powerful geometric observable for investigating the impact-parameter dependence of various initial and final state effects in \(eA\) collisions.Following the EIC Early Science Matrix, simulated event samples are generated with BeAGLE++ for \(e\)+Ag collisions at a center-of-mass beam configuration of \(9 \times 115~\mathrm{GeV}\), covering the kinematic range 	$1~\mathrm{GeV}^2 < Q^2 < 100~\mathrm{GeV}^2$. These simulated events are subsequently used to investigate observables sensitive to the collision geometry. Nuclear de-excitation following the primary electron--nucleus interaction proceeds predominantly through the evaporation of nucleons. Among the emitted particles, neutrons constitute the dominant component, followed by protons. This hierarchy arises because protons must overcome the nuclear Coulomb barrier before escaping the residual nucleus, whereas neutrons are not subject to Coulomb repulsion and therefore have a significantly higher emission probability. Consequently, evaporated neutrons provide the most direct probe of the nuclear evaporation process and serve as a key experimental observable for constraining the collision geometry in \(eA\) interactions.

In central collisions, corresponding to small impact parameters, the virtual photon interacts with nucleons deep inside the nucleus, resulting in a larger effective interaction length, \(d\). Consequently, the produced hadrons traverse a longer distance through the nuclear medium before exiting the nucleus, leading to enhanced final-state interactions and greater energy deposition into the residual nucleus. The increased nuclear excitation subsequently results in a higher probability of nucleon evaporation, predominantly in the form of neutrons, thereby producing a larger energy deposition in the ZDC. Conversely, peripheral collisions are characterized by shorter in-medium path lengths, reduced nuclear excitation, fewer evaporated neutrons, and correspondingly lower ZDC energy deposition. Figure~\ref{EZDC}(a) shows the generated-level distribution of the neutron energy deposited in the ZDC, \(E_{\rm n}^{\rm ZDC}\). The green shaded region corresponds to the most central \(0\text{--}1\%\) of events, defined as the top \(1\%\) of the event sample with the highest ZDC energy deposition (\(E_{\rm n}^{\rm ZDC}>1.59~\mathrm{TeV}\)). In contrast, the yellow shaded region represents the most peripheral \(60\text{--}100\%\) of events, characterized by \(E_{\rm n}^{\rm ZDC}<0.24~\mathrm{TeV}\). The correlation between the effective interaction length and the ZDC energy percentile is presented in Fig.~\ref{EZDC}(b), where the average interaction length, \(\langle d \rangle\), is plotted as a function of the ZDC energy percentile class. For the minimum-bias event sample (\(0\text{--}100\%\)), the average interaction length is found to be \(\langle d \rangle = 3.681~\mathrm{fm}\). A pronounced decrease in \(\langle d \rangle\) is observed when moving from the most central (\(0\text{--}1\%\)) to more inclusive centrality classes (e.g., \(0\text{--}10\%\)), confirming the strong sensitivity of the ZDC energy to the collision geometry in central interactions. In contrast, the variation of \(\langle d \rangle\) becomes relatively weak in the peripheral region, reflecting the reduced dependence of the interaction geometry on the impact parameter. Based on these observations, the \(0\text{--}1\%\) and \(60\text{--}100\%\) ZDC energy classes are adopted throughout this work to represent the most central and most peripheral collision samples, respectively. These results are consistent with the legacy BeAGLE simulations of \(e\)Pb collisions at a beam-energy configuration of \(18 \times 110~\mathrm{GeV}\) reported in Ref.~\cite{Chang:2022jgo}. The observed agreement demonstrates that the correlation between forward neutron production and collision geometry is robust against significant modifications to the underlying physics model. In particular, despite the implementation of a substantially revised treatment of final-state interactions (FSI) and nuclear de-excitation in BeAGLE++, the same geometric dependence is reproduced, providing strong validation of the method

Figure~\ref{centrality} presents the centrality determination for $e$Ag collisions based on the energy deposited in the Zero Degree Calorimeter (ZDC), as predicted by the BeAGLE++ simulations. The correlations between the deposited ZDC energy, $E_{\rm ZDC}$, and the collision geometry observables—the impact parameter, $b$, and the normalized nuclear thickness, $T(b)/\rho_{0}$—are shown in Fig.~\ref{centrality}(a) and (c), respectively. An increase in $E_{\rm ZDC}$ corresponds to a decrease in the average impact parameter and a simultaneous increase in the average nuclear thickness traversed by the virtual photon, reflecting the transition from peripheral to increasingly central interactions.

These correlations confirm the  sensitivity of forward neutron production to the underlying collision geometry. The corresponding normalized distributions of $b$ and $T(b)/\rho_{0}$ for the $0$--$1\%$ (most central) and $60$--$100\%$ (most peripheral) centrality classes are shown in Fig.~\ref{centrality}(b) and (d), where the distributions are normalized to the total number of generated events. A pronounced separation between the central and peripheral event classes is observed for both geometric quantities, demonstrating that the ZDC-based centrality estimator provides an effective means of selecting events with distinct impact parameters and nuclear path lengths. The correlations between $E_{\rm ZDC}$ and the geometric observables, as well as the resulting centrality distributions, are found to be  consistent with those obtained using the legacy BeAGLE framework, demonstrating that the new BeAGLE++ framework preserves the established centrality reconstruction while incorporating a more sophisticated treatment of the underlying physics.

\clearpage
\end{document}